\documentclass[11pt]{article}
\input epsf
\usepackage{psfrag}
\usepackage{graphicx}

\catcode`\@=11
\def\@citex[#1]#2{%
\if@filesw \immediate \write \@auxout {\string \citation {#2}}\fi
\@tempcntb\m@ne \let\@h@ld\relax \def\@citea{}%
\@cite{%
  \@for \@citeb:=#2\do {%
    \@ifundefined {b@\@citeb}%
      {\@h@ld\@citea\@tempcntb\m@ne{\bf ?}%
      \@warning {Citation `\@citeb ' on page \thepage \space undefined}}%
      {\@tempcnta\@tempcntb \advance\@tempcnta\@ne%
      \@tempcntb\number\csname b@\@citeb \endcsname \relax%
      \ifnum\@tempcnta=\@tempcntb 
        \ifx\@h@ld\relax%
          \edef \@h@ld{\@citea\csname b@\@citeb\endcsname}%
        \else%
          \edef\@h@ld{\ifmmode{-}\else--\fi\csname b@\@citeb\endcsname}%
        \fi%
      \else
        \@h@ld\@citea\csname b@\@citeb \endcsname%
        \let\@h@ld\relax%
      \fi}%
    \def\@citea{,\penalty\@highpenalty\,}%
  }\@h@ld
}{#1}}

\def\@citeb#1#2{{[#1]\if@tempswa , #2\fi}}
\def\@citeu#1#2{{$^{#1}$\if@tempswa , #2\fi }}
\def\@citep#1#2{{#1\if@tempswa , #2\fi}}

\def\bcites{         
        \catcode`\@=11
        \let\@cite=\@citeb
        \catcode`\@=12
}

\def\upcites{         
        \catcode`\@=11
        \let\@cite=\@citeu
        \catcode`\@=12
}

\def\plaincites{      
        \catcode`\@=11
        \let\@cite=\@citep
        \catcode`\@=12
}

\newcount\hour
\newcount\minute
\newtoks\amorpm
\hour=\time\divide\hour by 60
\minute=\time{\multiply\hour by 60 \global\advance\minute by-\hour}
\edef\standardtime{{\ifnum\hour<12 \global\amorpm={am}%
        \else\global\amorpm={pm}\advance\hour by-12 \fi
        \ifnum\hour=0 \hour=12 \fi
        \number\hour:\ifnum\minute<10 0\fi\number\minute\the\amorpm}}
\edef\militarytime{\number\hour:\ifnum\minute<10 0\fi\number\minute}

\def\draftlabel#1{{\@bsphack\if@filesw {\let\thepage\relax
   \xdef\@gtempa{\write\@auxout{\string
      \newlabel{#1}{{\@currentlabel}{\thepage}}}}}\@gtempa
   \if@nobreak \ifvmode\nobreak\fi\fi\fi\@esphack}
        \gdef\@eqnlabel{#1}}
\def\@eqnlabel{}
\def\@vacuum{}
\def\marginnote#1{}
\def\draftmarginnote#1{\marginpar{\raggedright\scriptsize\tt#1}}
\def\draft{
        \pagestyle{plain}
        \overfullrule=2pt
        \oddsidemargin -.5truein
        \def\@oddhead{\sl \phantom{\today\quad\militarytime} \hfil
        \smash{\Large\sl DRAFT} \hfil \today\quad\militarytime}
        \let\@evenhead\@oddhead
        \let\label=\draftlabel
        \let\marginnote=\draftmarginnote
        \def\ps@empty{\let\@mkboth\@gobbletwo
        \def\@oddfoot{\hfil \smash{\Large\sl DRAFT} \hfil}
        \let\@evenfoot\@oddhead}
        \def\@eqnnum{(\theequation)\rlap{\kern\marginparsep\tt\@eqnlabel}%
        \global\let\@eqnlabel\@vacuum}  }

\def\blackfonts{
        \font\blackboard=msbm10 scaled\magstep1
        \font\blackboards=msbm8
        \font\blackboardss=msbm6
}

\def\prep{         
        \catcode`\@=11
        \input art10.sty
        \catcode`\@=12
        
        \let\small\null
        \def\blackfonts{
                \font\blackboard=msbm10
                \font\blackboards=msbm7
                \font\blackboardss=msbm5
        }
        \let\sl\it
        \twocolumn
        \sloppy
        \voffset=-2.54truecm
        \hoffset=-2.54truecm
        \flushbottom
        \parindent 1em
        \leftmargini 2em
        \leftmarginv .5em
        \leftmarginvi .5em
        \marginparwidth 48pt
        \marginparsep 10pt
        \setlength{\columnsep}{2truecm}
        \setlength{\textwidth}{25.4truecm}
        \setlength{\textheight}{17truecm}
        \baselineskip=16pt
        \oddsidemargin .18truein
        \evensidemargin .17truein
}

\def\eqalign#1{\null\,\vcenter{\openup\jot\m@th
  \ialign{\strut\hfil$\displaystyle{##}$&$\displaystyle{{}##}$\hfil
      \crcr#1\crcr}}\,}
\def\eqalignno#1{\displ@y \tabskip\centering
  \halign to\displaywidth{\hfil$\@lign\displaystyle{##}$\tabskip\z@skip
    &$\@lign\displaystyle{{}##}$\hfil\tabskip\centering
    &\llap{$\@lign##$}\tabskip\z@skip\crcr
    #1\crcr}}

\def\section{\@startsection {section}{1}{\z@}{3.ex plus 1ex minus
 .2ex}{2.ex plus .2ex}{\large\bf}}
\def\subsection{\@startsection{subsection}{2}{\z@}{2.75ex plus 1ex minus
 .2ex}{1.5ex plus .2ex}{\bf}}        

\def\appendix{{\newpage\section*{Appendix}}\let\appendix\section%
        {\setcounter{section}{0}
        \gdef\thesection{\Alph{section}}}\section}

\def\abstract{\if@twocolumn
\section*{Abstract}
\else 
\begin{center}
{\bf Abstract\vspace{-.5em}\vspace{0pt}}
\end{center}
\quotation
\fi}

\catcode`\@=12

\def\d{\partial}

\def\sqr#1#2{{\vcenter{\vbox{\hrule height.#2pt\hbox{\vrule width.#2pt 
height#1pt \kern#1pt \vrule width.#2pt}\hrule height.#2pt}}}}

\def\=d{\,{\buildrel\rm def\over =}\,}

\def\i3p{\p32\int d^3p}

\def\As{A\hbox to 1pt{\hss /}}
\def\np4{\int d^4p_1\cdots d^4p_{n-1}\, }

\def\nx4{\int d^4x_1\ldots d^4x_n\, }

\def\kon#1#2{\vbox{\halign{##&&##\cr
\lower4pt\hbox{$\scriptscriptstyle\vert$}\hrulefill &
\hrulefill\lower4pt\hbox{$\scriptscriptstyle\vert$}\cr $#1$&
$#2$\cr}}}

\def\konv#1#2#3{\hbox{\vrule height12pt depth-1pt}
\vbox{\hrule height12pt width#1cm depth-11.6pt}
\hbox{\vrule height6.5pt depth-0.5pt}
\vbox{\hrule height11pt width#2cm depth-10.6pt\kern5pt
      \hrule height6.5pt width#2cm depth-6.1pt}
\hbox{\vrule height12pt depth-1pt}
\vbox{\hrule height6.5pt width#3cm depth-6.1pt}
\hbox{\vrule height6.5pt depth-0.5pt}}
\def\konu#1#2#3{\hbox{\vrule height12pt depth-1pt}
\vbox{\hrule height1pt width#1cm depth-0.6pt}
\hbox{\vrule height12pt depth-6.5pt}
\vbox{\hrule height6pt width#2cm depth-5.6pt\kern5pt
      \hrule height1pt width#2cm depth-0.6pt}
\hbox{\vrule height12pt depth-6.5pt}
\vbox{\hrule height1pt width#3cm depth-0.6pt}
\hbox{\vrule height12pt depth-1pt}}

\def\konw#1#2#3{\hbox{\vrule height12pt depth-1pt}
\vbox{\hrule height12pt width#1cm depth-11.6pt}
\hbox{\vrule height6.5pt depth-0.5pt}
\vbox{\hrule height12pt width#2cm depth-11.6pt \kern5pt
      \hrule height6.5pt width#2cm depth-6.1pt}
\hbox{\vrule height6.5pt depth-0.5pt}
\vbox{\hrule height12pt width#3cm depth-11.6pt}
\hbox{\vrule height12pt depth-1pt}}

\def\i{{\rm int}}

\def\e{{\rm ext}}

\def\a{{\rm av}}

\def\m3{{\mu_1\mu_2\mu_3}}

\def\p{{(+)}}

\def\be{\begin{equation}}       \def\eq{\begin{equation}}
\def\ee{\end{equation}}         \def\eqe{\end{equation}}

\def\bea{\begin{eqnarray}}      \def\eqa{\begin{eqnarray}}
\def\ena{\end{eqnarray}}        \def\eea{\end{eqnarray}}
                                \def\eqae{\end{eqnarray}}

\def\ba{\begin{array}}
\def\ea{\end{array}}
\def\unit{1 \hskip-.3em \raise2pt\hbox{$ \scriptstyle |$ } }

\def\a{\alpha}

\def\d{\delta}
\def\e{\epsilon}           

\def\i{\iota}

\def\m{\mu}
\def\n{\nu}
  
\def\p{\pi}                
\def\t{\tau}

\def\D{\Delta}

\def\bop#1{\setbox0=\hbox{$#1M$}\mkern1.5mu
        \vbox{\hrule height0pt depth.04\ht0
        \hbox{\vrule width.04\ht0 height.9\ht0 \kern.9\ht0
        \vrule width.04\ht0}\hrule height.04\ht0}\mkern1.5mu}

\def\>{\rangle} 

\def\<{\langle} 
\def\Dsl{D \hskip-.6em \raise1pt\hbox{$ / $ } }

\def\sl#1{\rlap{\hbox{$\mskip 1 mu /$}}#1}
\def\leftrightarrowfill{$\mathsurround=0pt \mathord\leftarrow \mkern-6mu
       \cleaders\hbox{$\mkern-2mu \mathord- \mkern-2mu$}\hfill
       \mkern-6mu \mathord\rightarrow$}
\def\dvec#1{\vbox{\ialign{##\crcr
       \leftrightarrowfill\crcr\noalign{\kern-1pt\nointerlineskip}
       $\hfil\displaystyle{#1}\hfil$\crcr}}}          
\def\hook#1{{\vrule height#1pt width0.4pt depth0pt}}
\def\leftrighthookfill#1{$\mathsurround=0pt \mathord\hook#1
       \hrulefill\mathord\hook#1$}
\def\underhook#1{\vtop{\ialign{##\crcr                 
       $\hfil\displaystyle{#1}\hfil$\crcr
       \noalign{\kern-1pt\nointerlineskip\vskip2pt}
       \leftrighthookfill5\crcr}}}
\def\smallunderhook#1{\vtop{\ialign{##\crcr      
       $\hfil\scriptstyle{#1}\hfil$\crcr
       \noalign{\kern-1pt\nointerlineskip\vskip2pt}
       \leftrighthookfill3\crcr}}}

\def\sfrac#1#2{{\vphantom1\smash{\lower.5ex\hbox{\small$#1$}}\over
       \vphantom1\smash{\raise.4ex\hbox{\small$#2$}}}} 
\def\bfrac#1#2{{\vphantom1\smash{\lower.5ex\hbox{$#1$}}\over
       \vphantom1\smash{\raise.3ex\hbox{$#2$}}}}      
\def\afrac#1#2{{\vphantom1\smash{\lower.5ex\hbox{$#1$}}\over#2}}  
\def\on#1#2{{\buildrel{\mkern2.5mu#1\mkern-2.5mu}\over{#2}}}
\def\ddt#1{\on{\hbox{\LARGE .\kern-2pt.}}#1}             
\def\tdt#1{\on{\hbox{\LARGE .\kern-2pt.\kern-2pt.}}#1}   

\def\boxes#1{
       \newcount\num
       \num=1
       \newdimen\downsy
       \downsy=-1.5ex
       \mskip-2.8mu
       \bo
       \loop
       \ifnum\num<#1
       \llap{\raise\num\downsy\hbox{$\bo$}}
       \advance\num by1
       \repeat}
\def\boxup#1#2{\newcount\numup
       \numup=#1
       \advance\numup by-1
       \newdimen\upsy
       \upsy=.75ex
       \mskip2.8mu
       \raise\numup\upsy\hbox{$#2$}}

\newskip\humongous \humongous=0pt plus 1000pt minus 1000pt
\def\caja{\mathsurround=0pt}
\def\eqalign#1{\,\vcenter{\openup2\jot \caja
       \ialign{\strut \hfil$\displaystyle{##}$&$
       \displaystyle{{}##}$\hfil\crcr#1\crcr}}\,}
\newif\ifdtup

\def\1ov4{{1\over 4}}

\def\ddt{\dot{\t}}

\renewcommand{\a}{\alpha}

\renewcommand{\d}{\delta}
\newcommand{\rmd}{{\rm d}}

\newcommand{\beq}{\begin{equation}}
\newcommand{\eeq}{\end{equation}}
\newcommand{\tD}{\tilde{\Delta}}
\def\ba{\begin{eqnarray}}
\def\ea{\end{eqnarray}}

\newcommand{\5}{\tilde{5}}

\begin{document}



\null\vskip-24pt
\hfill KL-TH 00/02
\vskip-10pt
\hfill {\tt hep-th/0002154}
\vskip0.3truecm
\begin{center}
\vskip 3truecm
{\Large\bf
Aspects of the conformal Operator Product Expansion in  AdS/CFT
Correspondence 
}\\ 
\vskip 1.5truecm
{\large\bf
Laurent Hoffmann, 
 \footnote{email:{\tt hoffmann@physik.uni-kl.de}} Anastasios C. Petkou
   \footnote{email:{\tt petkou@physik.uni-kl.de}}  
and Werner R\" uhl \footnote{
email:{\tt ruehl@physik.uni-kl.de}}
}\\
\vskip 1truecm
{\it Department of Physics, Theoretical Physics\\
University of Kaiserslautern, Postfach 3049 \\
67653 Kaiserslautern, Germany}\\

\end{center}
\vskip 1truecm
\centerline{\bf Abstract}

We present a detailed analysis of a scalar conformal four-point function
obtained from AdS/CFT correspondence. We study the 
scalar exchange graphs on AdS$_{d+1}$ and discuss their analytic
properties. Using methods of conformal partial wave
analysis, we present a general procedure to study conformal four-point
functions in terms of exchanges of scalar
and tensor fields. The logarithmic terms in the four-point function
are connected to the anomalous dimensions of the exchanged fields. 
Comparison of the results from  AdS$_{d+1}$
graphs with the conformal partial wave analysis suggests a possible
general form for the operator product 
expansion of scalar fields in 
the boundary CFT$_d$.

\newpage

\section{Introduction}

The duality
between string or $M$-theory compactifications on AdS$_{d+1}$ and
$d$-dimensional superconformal gauge theories suggested by AdS/CFT
correspondence \cite{maldacena} has been the subject of intensive
research over the past couple of years (for a recent review see
\cite{oz}). Gradually, the emerging 
picture takes the form of the long-sought string/gauge
theory relationship \cite{polyakov}. Recently, in a minkowsian version
of the correspondence the $d$-dimensional
conformal field theory (CFT) has been discussed in the context of
local quantum field theory \cite{haag} defined on a   
standard (flat) 
compactified Minkowski space $M^{c}_{1,d-1}$. This space arises as the
boundary of the AdS$_{1,d}$ space-time. The isometry group of both
spaces is $SO(d,2)$ and the state space of the boundary CFT is related
to the state space of the bulk theory \cite{rehren}. 

Such a view of the AdS/CFT correspondence implies that the known local
structure of conformal field theory, (see for example \cite{fradkin}
and references therein), is connected to the local structure of the 
the field (or string) theory living on AdS. In particular, harmonic
analysis  on the isometry group $SO(d,2)$ (``conformal 
partial wave analysis'' CPWA), of $n$-point
functions of the boundary CFT should be valid. This is
equivalent to the existence of an operator product expansion (OPE) for
the boundary CFT. Such expansions are convergent in a topology defined
by the $n$-point functions on which they are applied (CPWA), or into which
they are inserted (OPE). 
Perhaps the most well-known application ground for CPWA and OPEs is
the (Euclidean) case $d=4$, 
when the
boundary  CFT  is the ${\cal N}=4$ SYM theory with gauge group
$SU(N)$. In that case, the large-$N$, large-$\lambda$ expansion
($\lambda=g_{YM}^{2}N$ with $g_{YM}$ being the gauge coupling),
corresponds to a perturbative form of the AdS 
theory in terms of the so-called ``Witten graphs'' \cite{maldacena}. 
Technical exploitations of the AdS/CFT correspondence are mainly based
on this graphical expansion \cite{viswa,freedman0}. 

Our aim in this work is to make a thorough investigation of a 
four-point function of scalar fields in the boundary CFT obtained
from a graphical 
expansion in AdS. We choose to work in general dimensions $d$
to ensure a broad applicability of our results. In Section 2 we set
the stage for our study 
by considering a theory on AdS with a single cubic local interaction
term. This may be viewed as the minimal AdS theory leading to a non-trivial
four-point function in the boundary. 
Locality arguments applied to the boundary
CFT require the analyticity of the AdS calculations. In Section 3 we
present the results of the AdS calculations in the direct and
the crossed channels. The
direct channel poses no analyticity problems. Complications arise in
the crossed channels where non-analytic terms might arise. We
prove that the possible non-analytic terms drop out by virtue of
highly non-trivial identities for generalized hypergeometric
functions, thus demonstrating that the corresponding CFT amplitudes
admit an OPE. We present a systematic harmonic analysis of CFT four-point
functions in Section 4 based on conformal exchange graphs 
for scalar and tensor fields. These graphs enable us to obtain the
general contribution of scalar and tensor fields in a four-point
function when the OPE is inserted in the direct channel. An important
 point of our analysis is the interpretation of the logarithmic terms
 which appear in four-point function calculations in terms of the
 anomalous dimensions of 
 the exchanged 
 scalar and tensor fields. This interpretation is by no means new as
 it has already been used by Symanzik \cite{symanzik0} and two of the
 present authors \cite{ruehl1,tassos1}. In
 Section 5
we combine the AdS results with the direct channel OPE. This allows
the recursive determination of the anomalous dimensions and couplings
of all scalar and tensor fields in the OPE.  Our results seem to draw a
clear picture for 
the OPE of two scalar fields of the boundary CFT; it contains
a) the full  contribution from the boundary conformal
field which corresponds 
to the AdS-field in the cubic bulk interaction \footnote{Our results
  differ in this point from the recent claims in \cite{dhoker}.} and b)
infinite towers of 
conformal scalar and tensor fields whose canonical dimensions
(e.g. the part of the dimension which does not depend on the
coupling), and
tensor rank are
simply related to the dimensions of the external fields in the
OPE. Finally, we summarize our results and
comment on possible 
extensions of our program in Section 6.

\section{General remarks}
  
It is well-known that CFT determines the form of two- and three-point
functions of general tensor fields up to constants (for a review see
\cite{tassos2}). Although these constants capture in general
non-trivial dynamical effects, four-point functions are the minimal
ones whose functional form depends in an essential way on the
dynamics. From conformal
invariance four-point functions are determined only up to a general
analytic function of two variables. Namely, 
consider the four-point function 
\beq
\label{4pt1}
\langle {\cal O}_1(x_1){\cal O}_3(x_3){\cal O}_2(x_2){\cal
  O}_4(x_4)\rangle\,,
\eeq
where the scalar fields ${\cal O}_i(x_i)$, $i=1,..,4$ have 
dimensions $\Delta_i$ respectively.  
Denoting $x_{ij} =x_i-x_j$ 
the four-point function (\ref{4pt1}) can be expanded in powers of
$x_{ij}^2$, determined solely from $\Delta_i$, and an analytic function
$F(u,v)$ of the two biharmonic ratios
\beq
\label{biharm}
u = \frac{x_{13}^2 x_{24}^2}{x_{12}^2 x_{34}^2} \,\,\,\,\,,\,\,\,\,\,\,
v = \frac{x_{14}^2 x_{23}^2}{x_{12}^2 x_{34}^2}\,.
\eeq
There exists then a physical real analyticity domain for $F(u,v)$ with
\cite{ruehl1} 
\beq
|1+u-v|\leq 2u^{\frac{1}{2}}\,,\,\,\,\,\,\, |1+v-u|\leq
2v^{\frac{1}{2}}\,,
\eeq
and a possible expansion is
\beq
\label{Fexp1}
F(u,v) = \sum_{n,m=0}^{\infty} \,\frac{u^n (1-v)^m}{n!m!}\, A_{nm}(\D_i;d)\,.
\eeq
An expansion such as (\ref{Fexp1}) is useful in the {\it direct channel limit}
\beq
\label{dchann}
x_1\rightarrow x_3\,,\,\,\,\, x_2\rightarrow x_4\,,\,\,\,\,
(u\rightarrow 0, v\rightarrow 1)\,,
\eeq
and it is obtained
from the calculation of conformal graphs
\cite{ruehl1,ruehl2,tassos1,freedman1,liu}.  In almost all
four-point function calculations one relies on a perturbative
expansion in some small parameter 
e.g. a coupling constant or $1/N$.  Such a perturbative expansion implies
the dependence of conformal dimensions on
the coupling constants (or $1/N$) and gives, for a specific
graph $\Gamma$, an expansion of the form
\beq
\label{Fexpgama}
F_{\Gamma}(u,v) = \sum_{n,m=0}^{\infty} \,\frac{u^n
  (1-v)^m}{n!m!}\left[\sum_{k=0}^{\cal{K}} A_{nm}^{(k)}(\D_i;d)\,(\ln u)^k
    \right]\,,
\eeq
where ${\cal K}$ depends on the perturbative order. One important
point here is the 
appearance of the logarithmic terms on the r.h.s. of
(\ref{Fexpgama}). These terms are not involved in the discussion of
the analytic properties of the expansion (\ref{Fexpgama}) as they can
in principle be summed up and exponentiated giving just an ``anomalous''
contribution to the dimensions of the exchanged fields
\cite{symanzik0,ruehl1,ruehl2,tassos1}. Logarithmic terms frequently
occur in conformal 
$n$-point functions and are in fact necessary in order to ensure the correct
conformal properties when the external points come close together
\cite{tassos2,tassos3}. 

In the context of AdS/CFT correspondence \cite{maldacena} one is
equipped with a standard procedure to generate a perturbative expansion
for conformal four-point functions. Namely, the relation of a (local)
field theory on AdS  to a CFT on the boundary can be
schematically written as  
\beq
Z_{{\rm AdS}}\equiv \int({\cal D}\phi)e^{-S[\phi]} \rightarrow
Z[\phi_0]\equiv \langle e^{\int\rmd^{d}x\,\phi_0(x)\,{\cal
    O}_{\phi}(x)}\rangle =e^{W_{{\rm CFT}}[\phi_0]}\,,
\eeq
where $W_{{\rm CFT}}[\phi_0]$ is the generating functional for connected
$n$-point functions of the field ${\cal O}_{\phi}(x)$ in the boundary
CFT, when $\phi_0(x)$ plays the role of an external source. Passing from $Z$ to
$Z[\phi_0]$ involves  solving the classical 
field equations for the AdS field \footnote{We consider the Euclidean
version  of 
AdS$_{d+1}$  space where  $\rmd \hat{x}^{\m}\rmd
\hat{x}_{\m}=\frac{1}{x_{0}^{2}} 
(\rmd x_{0}\rmd 
x_{0}+\rmd x^{i}\rmd x^{i})$, with $i=1,..,d.$ and
$\hat{x}_{\m}=(x_{0},x_{i})$. 
The boundary of this space is isomorphic
to {\bf{S}}$^{\rm d}$ since it consists of
{\bf{R}}$^{\rm d}$ at $x_{0}=0$ and a single point at $x_{0}=\infty$.}
$\phi(x_0,x)$ with boundary
conditions such that $\phi(x_0,x)|_{\partial{\rm AdS}} =
\phi_0(x)$. When the AdS action involves non-quadratic terms  it
can only be 
evaluated within a perturbative
expansion in the coupling constant. Such an expansion has a definite
interpretation in terms of ``Witten graphs''
\cite{maldacena,viswa,freedman0} connecting points in
the boundary of AdS via interactions taking place in the bulk. The
result is a perturbative expansion for the $n$-point functions
of the boundary CFT. 

Here we consider the simplest local field theory on
AdS which gives rise to non-trivial four-point functions of the
boundary CFT. Namely, we consider the following action for the AdS scalar
fields $\phi(\hat{x})$ and  $\sigma(\hat{x})$ 
\ba
\label{action}
S &=& \int \rmd^{d+1}\hat{x}\sqrt{g}
  \left(\sfrac{1}{2}\partial_{\m}\phi(\hat{x}) 
  \partial^{\m}\phi(\hat{x}) +\sfrac{1}{2}\tilde{m}^2\phi^{2}(\hat{x})+
  \sfrac{1}{2}\partial_{\m}\sigma(\hat{x}) 
  \partial^{\m}\sigma(\hat{x})
  +\sfrac{1}{2}m^2\sigma^{2}(\hat{x})\right. \nonumber \\ 
  & &\hspace{2cm}\left. +\sfrac{\gamma_*}{2}\phi^2(\hat{x})
    \sigma(\hat{x})\right)\,.
\ea
The action (\ref{action}) gives rise to a boundary CFT of the scalar
fields ${\cal 
  O}_{\phi}(x)$ and ${\cal O}_{\sigma}(x)$ with corresponding
two-point functions \cite{viswa,freedman0}
\ba
\label{2ptphi}
\langle{\cal O}_{\phi}(x_1){\cal O}_{\phi}(x_2)\rangle & = &
C_{\tilde{\Delta}} \frac{1}{x_{12}^{2\tilde{\Delta}}} \,,\,\,\,\,\,
\tilde{\Delta}=\sfrac{d}{2}+\sqrt{\tilde{m}^2+\sfrac{d^2}{4}} \,,\\
\langle{\cal O}_{\sigma}(x_1){\cal O}_{\sigma}(x_2)\rangle & = &
C_{\Delta} \frac{1}{x_{12}^{2\Delta}} \,,\,\,\,\,\, 
\Delta = \sfrac{d}{2}+\sqrt{m^2+\sfrac{d^2}{4}}\label{2ptsigma}\,,\\
C_{\D}& = &
\frac{2(\D-\frac{1}{2}d)\Gamma(\D)}{\pi^{\frac{1}{2}d}\Gamma(\D-\frac{1}{2}d)}
\,. 
\ea
It also implies the existence of the three-point function
\cite{viswa,freedman1} 
\ba
\label{3ptf}
\langle{\cal O}_{\phi}(x_1){\cal O}_{\phi}(x_2){\cal
  O}_{\sigma}(x_3)\rangle & = & \gamma_* \,
g_{\tilde{\Delta}\tilde{\Delta}\Delta}
\frac{1}{(x_{12}^2)^{\tilde{\Delta}-\frac{1}{2} \Delta}(x_{13}^2
  x_{23}^2)^{\frac{1}{2} \Delta}}\,, \\
g_{\tilde{\Delta}\tilde{\Delta}\Delta} & = &
\frac{1}{4\pi^d}\frac{\Gamma^2(\frac{1}{2}\Delta)
  \Gamma(\tilde{\Delta}-\frac{1}{2}\Delta)
  \Gamma(\tilde{\Delta}+\frac{1}{2} \Delta
  -\frac{1}{2}d)}{\Gamma^2(\tilde{\Delta} -\frac{1}{2}d)
  \Gamma(\Delta -\frac{1}{2}d)}\,.
\ea 
The coupling constant $\gamma_*$ is the parameter which induces the
non-trivial dynamics of the boundary CFT. In
principle, one should have information
regarding its magnitude before trying to make sense of the
``Witten graph'' expansion. This is the case when $\phi(\hat{x})$ and
$\sigma(\hat{x})$ correspond to
Kaluza-Klein modes of the compactified supergravity theory and $\gamma_*$ is 
determined by the standard reduction procedure
\cite{seiberg,gleb1} to be $\gamma_* \sim O(1/N)$, where
$SU(N)$ is the 
gauge group of the boundary CFT \footnote{Recently, the full 
AdS$_{d+1}$ action for the Kaluza-Klein modes has been evaluated up to
quartic couplings \cite{gleb2}.}. In this case one is able
to order the perturbative expansion of the action (\ref{action})
according to the  number of  
cubic vertices. For the purposes of our work it suffices to 
assume that $\gamma_*$ is also a ``small'' parameter when
$\phi(x)$ and 
$\sigma(x)$ are general scalar fields, as our results do not depend
in an essential way on its magnitude.

\begin{figure}[h]
\centering
\begin{minipage}{15cm}
\centering
\psfrag{1}{${}$}
\psfrag{x1}{$x_1$}
\psfrag{x2}{$x_2$}
\psfrag{x3}{$x_3$}
\psfrag{x4}{$x_4$}
\psfrag{D}{$A(x_1,x_3;x_2,x_4)\,=$}
\psfrag{C}{$B(x_1,x_3;x_2,x_4)\,=$}
\includegraphics[width=12cm]{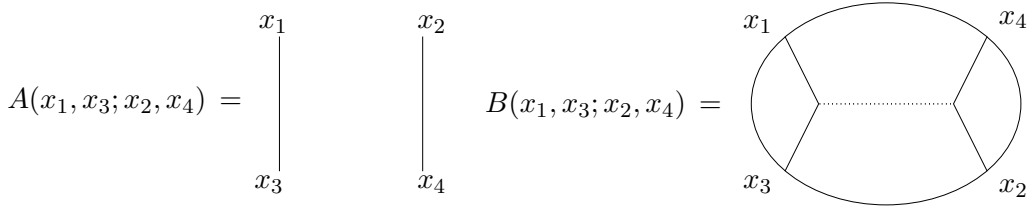}
\centering
\caption{\it \small The $A$ and $B$ graphs. In the $A$ graph the
solid lines correspond to the full ${\cal O}_{\phi}(x)$ propagator
(\ref{2ptphi}). In the $B$ graph, solid lines correspond to the
``bulk-to-boundary'' propagators (\ref{btbo}) and the dotted line to the
``bulk-to-bulk'' one (\ref{btbu})}.
\end{minipage}
\end{figure} 

Our main interest is in the four-point function of the scalar field
${\cal O}_{\phi}(x)$. Up to tree ``Witten graphs'' standard AdS/CFT
calculations give the following expansion  
\ba
\label{4pt2}
\langle {\cal O}_{\phi}(x_1){\cal O}_{\phi}(x_3) {\cal
  O}_{\phi}(x_2) {\cal O}_{\phi}(x_4)\rangle
&=& A(x_1,x_2;x_3,x_4) +A(x_1,x_3;x_2,x_4)  +
A(x_1,x_4;x_3,x_2) \nonumber\\ 
&  & \hspace{-1.8cm}+ \gamma_*^2\left[B(x_1,x_2;x_3,x_4) +
  B(x_1,x_3;x_2,x_4) + B(x_1,x_4;x_3,x_2)\right]\,, 
\ea 
where the $A$ and $B$ terms are depicted in Fig.1.  Note that we
consider the full four-point function 
and not only its connected part. 
The explicit expressions for the graphs $A$ and $B$ are given by
\ba
\label{agraps}
 A(x_1,x_2;x_3,x_4) & = &
 C^2_{\tilde{\Delta}} \frac{1}{(x_{12}^2 x_{34}^2)^{\tilde{\Delta}}}\,,\\
\label{bgraphs}
B(x_1,x_2;x_3,x_4) & = &
\int\frac{\rmd^{d+1}\hat{y}\,\rmd^{d+1}\hat{z}}{y_0^{d+1}z_0^{d+1}}
K_{\tilde{\Delta}}(x_1,\hat{y})K_{\tilde{\Delta}}(x_2,\hat{y})
G_{\Delta}(\hat{y},\hat{z}) K_{\tilde{\Delta}}(x_3,\hat{z})
K_{\tilde{\Delta}}(x_4,\hat{z})\,,  
\ea
and the standard forms of the ``bulk-to-boundary'' and ``bulk-to-bulk''
propagators that we use are \cite{freedman1,liu}
\ba
\label{btbo}
K_{\Delta}(x,\hat{y}) & = & k_{\Delta}
\left[\frac{x_0}{x_0^2 + (x-y)^2}\right]^{\Delta} \,,\,\,\,\,\,\,\,\,
k_{\Delta} \,\,=\,\,\frac{\Gamma(\Delta)}{\pi^{\frac{1}{2}d} 
  \Gamma(\Delta -\frac{1}{2}d)}\,,\\ 
\label{btbu}
G_{\Delta}(x,y) & = & {\cal G}_{\Delta}\, q^{-\Delta}{}_{2}F_{1}\left(
  \sfrac{1}{2}(\Delta 
  +1), \sfrac{1}{2}\Delta; \Delta-\sfrac{d}{2}+1; q^{-2}\right)\,, \\
{\cal G}_{\Delta} & = & \frac{\Gamma(\Delta)}{ 2^{\Delta
    +1}\pi^{\frac{1}{2}d}\Gamma(\Delta -\frac{1}{2}d +1)}
\,\,\,\,,\,\,\,\,\, q^2 \,\,=\,\, \frac{x_0^2 +y_0^2 +(x-y)^2}{2x_0 y_0}\,.
\ea
The disconnected $A$ graphs simply give
\beq
\label{disc}
\frac{C^2_{\tD}}{(x_{13}^2x_{24}^2)^{\tilde{\Delta}}} +
\frac{C^2_{\tD}}{(x_{12}^2x_{34}^2)^{\tilde{\Delta}}}
\left[1+v^{-\tilde{\Delta}}\right]\,. 
\eeq
The integrations in the ``exchange graphs'' $B$ can be done
using either Symanzik's method \cite{symanzik} or following \cite{liu}
(see Appendix A), when one obtains the
following convenient 
representation  as a Mellin-Barnes integral
\ba
\label{mb1}
B(x_1,x_3;x_2,x_4) & = & \frac{\kappa}{(x_{12}^2x_{34}^2)^{\tD}}
\int_{{\cal C}}\frac{\rmd s}{2\pi{\rm i}} \Gamma^2(-s) \Biggl[
\frac{\Gamma^4(\tD+s) \Gamma^2(\frac12 \D+\tD-\frac12 d)
  \Gamma(\frac12 \D-\tD-s)}{ \Gamma(2\tD +2s)\Gamma(\D-\frac12
  d+1)\Gamma(\frac12 \D+\tD-\frac12 d-s)} \nonumber \\
&& \hspace{-2.5cm} \times \,_3F_2\left( \sfrac12 \D+\tD-\sfrac12
  d,\sfrac12 \D+\tD-\sfrac12 
  d,\sfrac12 \D-\tD-s; \D-\sfrac12 d +1,\tD+\sfrac12\D-\sfrac12
  d-s;1\right) \nonumber \\
& & \hspace{3cm}\times \,u^s \,_2F_1\left(\tD +s,\tD +s;2\tD +2s;
  1-v\right)\Biggl]\,, 
\ea  
where 
\beq
\kappa =
\frac{\Gamma(2\tilde{\Delta} -\frac{1}{2}d)}{8\pi^{3d/2}\Gamma^4(\tilde{\Delta}
-\frac{1}{2}d)}\,. 
\eeq
The remaining two crossing symmetric $B$ graphs are obtained from
(\ref{mb1}) by suitable interchanges of the $x$'s. 
Our strategy is now to present an explicit expression for the
four-point function (\ref{4pt2}) and study its analyticity properties.
This expression will then be compared with the
CFT expectation which is simply the CPWA of the four-point function
(\ref{4pt2}). The result will suggest the local form for the OPE of the
conformal scalar field ${\cal O}_{\phi}(x)$ with itself.

\section{Explicit results for the AdS exchange graphs}

In this section we present the results for the AdS ``exchange graphs''
and discuss their analyticity properties. 

\subsection{The direct channel}

Our starting point is the Mellin-Barnes representation (\ref{mb1}) for
the $B$ terms in (\ref{4pt2}). It is easy to see that
(\ref{mb1}) is suitable for studying the direct channel limit
(\ref{dchann}). The result of the integration is a double series
expansion coming from the summation over the $\Gamma$-function poles
included in the contour ${\cal C}$. It can be written down as (see
Appendix A for the details) 
\beq
\label{b1}
B(x_1,x_3;x_2,x_4) = \frac{1}{(x_{12}^2
  x_{34}^2)^{\tilde{\Delta}}}
  \sum_{m,n=0}^{\infty}\frac{u^n(1-v)^m}{n!m!}\left[ -a^{(1)}_{nm}\ln u
    + b^{(1)}_{nm} +u^{\frac12\D -\tD}c^{(1)}_{nm}\right]\,,
\eeq
where
\bea
\label{b1a}
a^{(1)}_{nm} & = & -\kappa\frac{\Gamma^2(\tD+n)\Gamma^2(\tD+n+m)}{
  (n+1)!\Gamma(2\tD+2n+m)}\frac{1}{(\frac12\D+\tD-\frac12
  d)}\nonumber \\
& & \times\,_3F_2(\sfrac12\D-\tD+1,\sfrac12\D+\tD-\sfrac12 d+1+n,1;n+2,
  \sfrac12\D+\tD- \sfrac12 d+1;1)\,,\\
\label{b1b}
b^{(1)}_{nm} & = & -\kappa\frac{\Gamma^2(\tD+n)\Gamma^2(\tD+n+m)}{
  \Gamma(2\tD+2n+m)} \Biggl\{ \frac{\Gamma(
  \tD-\frac12\D)\Gamma(\tD+\frac12\D-\frac12 d)\,(n)!}{\Gamma(2\tD-\frac12
  d)(\tD-\frac12\D)_{n+1}(\tD+\frac12\D-\frac12 d)_{n+1}} \nonumber \\
& & +\frac{\psi(\frac12\D+\tD-\frac12 d+n+1)-2\psi(\tD+n)-
  2\psi(\tD+n+m)+2\psi(2\tD+2n+m)
  }{(n+1)!\,(\frac12\D+\tD-\frac12 
  d)}
\nonumber \\
& & 
\hspace{.5cm} \times\,_3F_2(\sfrac12\D-\tD+1,\sfrac12\D+\tD-\sfrac12
d+1+n,1;n+2,  
  \sfrac12\D+\tD- \sfrac12 d+1;1)\nonumber \\
& & +\frac{1}{(\frac12\D+\tD-\frac12
  d)_{n+1}} \sum_{r=0}^{\infty}\frac{(-n)_r(1+\frac12
  d-\frac12\D-\tD)_r}{(r!)^2(\tD-\frac12\D+r)}\,\psi(n+1-r)\Biggl\}\,,\\
\label{b1c}
c^{(1)}_{nm} & =
&\kappa\,\frac{\Gamma^4(\frac{1}{2}\Delta)
  \Gamma^2(\tilde{\Delta} -\frac{1}{2}\Delta)\Gamma^2(\tilde{\Delta}
  +\frac{1}{2}\Delta -\frac{1}{2}d)}{\Gamma(\Delta)
  \Gamma(2\tilde{\Delta}-\frac{1}{2}d) \Gamma(\Delta-\frac{1}{2}d+1)}
\frac{(\frac{1}{2}\Delta)^2_{n}(\frac{1}{2}\Delta)^2_{n+
    m}}{(\Delta)_{2n+m}(\Delta-\frac{1}{2}d+1)_{n}}\,. 
\ea
The Pochhammer symbol $(a)_n$ is defined as
\beq
(a)_n = \frac{\Gamma(a+n)}{\Gamma(a)}\,.
\eeq
The hypergeometric functions which appear in (\ref{b1a}) and
(\ref{b1b}) can be given in term of terminating series by virtue of
the identity given in (\ref{ap4}). However, due to 
identities of the form
\beq
\label{psiterm}
\lim_{a\rightarrow
0}\frac{\psi(a-r)}{\Gamma(a-r)}=(-1)^{r+1}\Gamma(r+1)
\eeq
the last term in (\ref{b1b}) gives a contribution proportional to a
$_4F_3$ generalized hypergeometric function.  
It is easy to see that (\ref{b1}) is analytic in the direct channel
limit $u\rightarrow 0$ and $v\rightarrow 1$.

\subsection{The crossed channel}
 
The calculations in both the crossed channels ($x_3\leftrightarrow x_4$
or $u\leftrightarrow v$ and $x_3\leftrightarrow x_2$ or $u\leftrightarrow
1/u$ and $v\leftrightarrow v/u$),  are significantly more
complicated as they involve the analytic continuation of the result
(\ref{b1}). The reason is that we want to obtain an expression which
can be matched with the {\it direct channel} OPE, therefore we require
that the result be written e.g. in the general  form of
(\ref{Fexpgama}). Consider for clarity the crossed channel obtained
from (\ref{mb1}) by the interchange $u\leftrightarrow v$. 

This can be achieved in two fashions. We start first from the
Mellin-Branes integral for $B(x_1,x_4;x_2,x_3)$ (say) analogous to
(\ref{mb1}), and expand it into contributions of the poles in $s$ to the
right of the integral contour. We obtain a decomposition into a
contribution from $\mathcal{O}_{\sigma}$ exchange and from the
exchange of an infinite tower of the tensor fields. Because of the
missing shadow terms (see Sections 4.1, 4.2) each contribution is
singular at $v = 1$. In
Appendix B we show that for the specific case of the B graph in
Fig. 1, the non-analytic terms cancel each other by virtue of highly
non-trivial identities for the generalized hypergeometric function
$_3F_2$. The explicit form of the remaining analytic terms can also be
derived using the generalized hypergeometric differential equation and
the representations of its solutions by Mellin-Barnes integrals
(Appendix B).\\ The second method  makes use of the Mellin-Barnes
integral for $B(x_1,x_4;x_2,x_3)$ itself and we observe that (after
the exchange of $u \leftrightarrow v$ in (21)) the power $v^s$ can be
Taylor expanded at $v = 1$ under the integral sign (Appendix B).
Here we just give a general formula which is
useful in extracting information for the possible general structure of
the conformal OPE from AdS/CFT correspondence. Namely, the result for
both the crossed channel graphs obtained from Fig.1 is of the form
\beq
\label{b2b3}
B(x_1,x_4;x_2,x_3) + B(x_1,x_2;x_3,x_4) = \frac{1}{
  (x_{12}^2  x_{34}^2)^{\tD}} \sum_{n,m=0}^{\infty}
\frac{u^n(1-v)^m}{n!m!} \left[-\tilde{a}_{nm}\ln
  u+\tilde{b}_{nm}\right] \,,
\eeq
where the coefficients $\tilde{a}_{nm}$ ,
$\tilde{b}_{nm}$  depend solely on $\D$, $\tD$  and $d$.

\section{General scalar and tensor exchange in CFT}

In order to analyze the AdS results of Section 3 in terms of CFT
partial waves, we need a formalism which allows the identification of
a general conformal tensor and all its derivatives in the four-point
function (\ref{4pt2}). To achieve this we construct four-point
amplitudes with covariant vertices and a general tensor field of rank
$l$ and dimension $\Delta$ exchanged in the direct channel. The
appropriate conformal graph is depicted in Fig.2  and gives the
conformally invariant local result
$\beta_{\tilde{\Delta}}(x_1,x_2,x_3,x_4;\Delta,l)$. 

\begin{figure}[h]
\centering
\begin{minipage}{15cm}
\centering
\psfrag{x1}{$x_1$}
\psfrag{x2}{$x_2$}
\psfrag{x3}{$x_3$}
\psfrag{x4}{$x_4$}
\psfrag{bb}{$\beta_{\tD}(x_1,x_3;x_2,x_4;\D,l)\,=$}
\psfrag{dl}{$\D\,\,,\,\,l$}
\includegraphics[width=7cm]{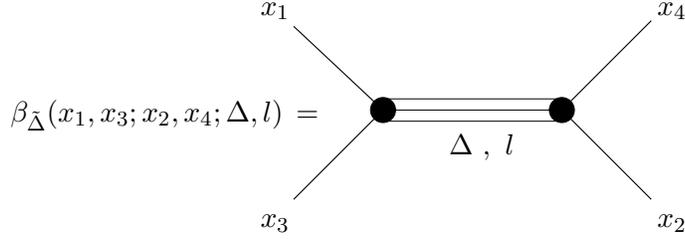}
\centering
\caption{\it \small The tensor exchange graphs. The dark blobs
correspond to the full vertex functions obtained by suitable
amputation of (\ref{mvertex}).}
\end{minipage}
\end{figure} 

\subsection{Scalar exchange in CFT}

To begin with we consider the exchange of a general scalar field of
dimension $\Delta$ in CFT. This corresponds to the $l=0$ graphs in
Fig.2. The three-point functions (\ref{3ptf}) which appear in it are
contracted with the inverse two-point function. The latter is obtained
from the two-point function (\ref{2ptsigma}) as
\ba
& &\int\rmd^d z\langle{\cal O}_{\sigma}(x) {\cal O}_{\sigma}(z)\rangle
\left[\langle {\cal
  O}_{\sigma}(z){\cal O}_{\sigma}(y)\rangle\right]^{-1} \,=\,
\delta^d(x-y) \,,\\
& & \left[\langle {\cal
  O}_{\sigma}(x){\cal O}_{\sigma}(y)\rangle\right]^{-1} \,=\,
\frac{1}{C_{\D}} \frac{\a(\D-\frac12
  d)}{\pi^d\a(\D)}\frac{1}{(z-y)^{2(d-\D)}} \,,\,\,\,\a(b)
\,=\, \frac{\Gamma(\frac12 d-b)}{\Gamma(b)}\,.
\ea
Then, we can use the D'EPP
formula \cite{depp} 
\bea
\int\rmd^d x\frac{1}{(x_1-x)^{2a_1}(x_2-x)^{2a_2}(x_3-x)^{2a_3}} &=&
 \frac{U(a_1,a_2,a_3)}{(x_{12}^2)^{\frac{1}{2}d-a_3}
  (x_{13}^2)^{\frac{1}{2}d-a_2}(x_{23}^2)^{\frac{1}{2}d-a_1}}\,,\\
U(a_1,a_2,a_3) & = & \pi^{\frac{1}{2}d}\a(a_1)
  \a(a_2)
  \a(a_3) 
\ea
which is valid only for $a_1+a_2+a_3=d$, to obtain for the scalar exchange
graph
\ba
\label{4star}
\beta_{\tilde{\Delta}}(x_1,x_3,x_2,x_4;\Delta) & = & \frac{(\gamma_*
  g_{\tD\tD\D})^2}{C_{\tD}} \frac{\Gamma(\D)\Gamma^2(\frac12 d-\frac12
  \D)}{\pi^{\frac12 d}\Gamma(\frac12
  d-\D)\Gamma^2(\frac12\D)}\frac{1}{(x_{13}^2)^{\tD-\frac12
    \D}(x_{24}^2)^{\tD+\frac12 \D-\frac12 d}} \nonumber\\
& & \times \int\rmd^d x \frac{1}{[(x_1-x)^{2}(x_3-x)^{2}]^{\frac12\D}
[(x_2-x)^{2} 
  (x_4-x)^{2}]^{\frac12 d-\frac12\D}}\,. 
\ea
The integral in (\ref{4star}) falls into the class of conformal 4-star
integrals \cite{ruehl1,ruehl2,tassos1} which can be evaluated using
Symanzik's technique \cite{symanzik} due to the {\it uniqueness}
condition satisfied by the massless scalar propagators involved in it
i.e. the sum of the dimensions of the propagators is $d$. The result
of the final integration is \cite{ruehl1,tassos1}
\ba
\label{cftex} \hspace{-.5cm}\beta_{\tD}(x_1,x_3;x_2,x_4;\D) & =
&\frac{b\,u^{-\tD}}{(x_{12}^2x_{34}^2)^{\tD}}\sum_{n,m=0}^{\infty}
\frac{u^n(1-v)^m}{n!m!}\Bigl[u^{\frac{1}{2}\D}c_{nm}(\D)
+u^{\frac{1}{2}d-\frac 
  {1}{2}\D} c_{nm}(d-\D)\Bigl],  \\
c_{nm}(\D)&= & \a(\D)\a^2(\sfrac12 d- \sfrac12\D)
\frac{(\frac{1}{2}\D)_n^2
  (\frac{1}{2}\D)_{n+m}^2}{(\D)_{2n+m}(\D-\frac{1}{2}d+1)_n}\,,\\
b & = & \frac{(\gamma_* g_{\tD\tD\D})^2}{C_{\D}} \frac{\a^2(\frac12\D)}{
  \a(\D)}\,. 
\ea
The first term on the r.h.s. of  (\ref{cftex}) is
the full contribution 
of a scalar field with dimension $\D$. Its form is fully determined by
conformal invariance \cite{parisi,fradkin,todorov,ruehl1,tassos1} and
involves an infinite number of descendants of the 
relevant scalar
field. One important observation here is that the second term in (\ref{cftex})
is obtained from the first by the replacement
\beq
\label{shad}
\Delta \rightarrow d-\Delta\,.
\eeq
This second infinite series in (\ref{cftex}) represents the so-called
{\it shadow symmetric} 
  singularities of the first series. \footnote{The term {\it shadow
    symmetry} was 
  introduced for the first time in \cite{parisi}. It corresponds to
  an intertwiner \cite{koller} of the conformal group in $d>2$ that
  maps the equivalent representations with dimensions $\eta$ and
  $d-\eta$ onto each other. Shadow symmetric singularities
  may correspond to physical {\it shadow fields} if the dimensions of
  the latter satisfy
  the unitarity bound e.g. $d-\eta\geq d/2-1$. See 
  \cite{ruehl1,ruehl2,tassos1} and also \cite{klebanov}.} The appearance of
  the shadow singularities is  
necessary for the cancellation of the non-analytic terms in the
crossed channel of standard CFT exchange graphs \cite{HPR1}. The
absence of shadow singularities in AdS calculations was some kind of a
puzzle and its solution was proposed in the use of irregular boundary
conditions \cite{klebanov,viswa2}. The correct solution, however,  is simple
and physically 
interesting. The holographic image of AdS supergravity and all the fields
which can be produced from it by operator product expansions are
gauge invariant composite fields (synonymously: conformal normal
products) of a set of basic fields, which we guess to be the vector
supermultiplet of SYM field theory. These basic fields are not contained
in the holographic image. On the other hand composite fields appear only
in operator pruduct expansions which are convergent power series with
increasing powers of the small distance. Shadow terms would therefore lead
to a series with decreasing powers and would make the whole series
twosided. There would not be a maximal small distance singularity.
For more details see \cite{LeonRuehl}.

\subsection{Tensor exchange in CFT}

The exchange of traceless symmetric tensors of dimension $\Delta$ and
rank $l$, corresponding to irreducible representations of dimension
$\D$ and spin $l$ of $SO(d,2)$,  can be also
calculated in CFT as the relevant graphs 
reduce to sums of scalar exchanges. For this we need to know the
general expression for the conformally invariant three-point function
of a symmetric, traceless tensor with dimension $\D$ and rank $l$ with
two scalar field of dimension $\tD$. This is determined from conformal
invariance up to an overall constant $g_{\tD,\tD,\D,l}$
\cite{fradkin,todorov2}. We use the vectors 
\ba
\label{ksi1}
\xi_{\mu}(1,2;3) & = & \frac{(x_{13})_{\mu}}{x_{13}^2}
-\frac{(x_{23})_{\mu}}{x_{23}^2} \,,\\
\label{ksi2}
\xi^2(1,2;3) & = & \frac{x_{12}^2}{x_{13}^2 x_{23}^2}\,,
\ea
to express the three-point function as \cite{todorov,fradkin}
\ba
\label{mvertex}
& &\hspace{-1cm} \langle{\cal O}(x_1){\cal
O}(x_3)M_{\mu_1,\mu_2,..,\mu_l}(x_5)\rangle 
= \frac{g_{\tilde{\Delta}\tilde{\Delta}
    \Delta,l}\,\,{\cal N}(\tD;\D,l)
  }{(x^2_{13})^{\tilde{\Delta}-\frac{1}{2}\Delta}(x_{15}^2  
  x_{35}^2)^{\frac{1}{2}\Delta}} \left[
  \frac{\xi_{\mu_1}\xi_{\mu_2}\cdots
    \xi_{\mu_l}}{(\xi^2)^{\frac{1}{2}l}} -{\rm trace
    \,\,terms}\right]\,,\\
&&\hspace{-1cm} {\cal N}(\tD;\D,l) = \frac{2^{\tD+\frac12\D+\frac12
    l}}{(2\pi)^{\frac12 d}} \left(\frac{\Gamma(\tD+\frac12\D+\frac12
    l-\frac12 d) \Gamma(\tD-\frac12\D+\frac12
    l)\Gamma^2(\frac12\D+\frac12 l)}{ \Gamma(d-\tD-\frac12\D+\frac12
    l)\Gamma(\frac12 d-\tD+\frac12\D+\frac12 l)\Gamma^2(\frac12
    d-\frac12\D+\frac12 l)}\right)^{\frac12}
\ea
where $\xi_{\mu}\equiv \xi_{\mu}(1,3;5)$. Then, the general 
amplitude depicted in 
Fig.2 consists of two covariant vertex functions one of which
is amputated and can be written as
\ba
\label{beta}
\beta_{\tilde{\Delta}}(x_1,x_2,x_3,x_4;\Delta,l) & = & \int\rmd^{d}x_{\5}\Bigl[
\sum_{\mu_1,..,\mu_l}\langle {\cal O}(x_1){\cal O}(x_3)
M_{\mu_1,\mu_2,..\mu_l}(x_{\5})\rangle \nonumber \\
& & \hspace{2.5cm}\langle M_{\mu_1,\mu_2,..,\mu_l}(x_{\5}){\cal O}(x_2)
{\cal O}(x_4)\rangle_{{\rm amp}}\Bigl]\,.
\ea 
The amputation of the second vertex in (\ref{beta}) is done on the
tensor field $M_{\mu_1,..,\mu_l}(x)$. With the following normalization
for the two-point function \cite{todorov2,fradkin}
\ba
\label{t2ptf}
& &\langle M_{\m_1,..,\m_l}(x_1)M_{\n_1,..,\n_l}(x_2)\rangle =
C_{\D,l}\frac{{\cal N}(\D,l)}{x_{12}^{2\D}}
\left[\Bigl\{I_{\m_1\n_1}(x_{12}) \cdots
    I_{\m_l\n_l}(x_{12})\Bigl\}_{sym}-{\rm traces} \right]\,,\\
&&{\cal N}(\D,l) = \frac{2^{\D}\Gamma(\D+l)\Gamma(d-\D-1)}{(2\pi)^{\frac12
    d}\Gamma(\frac12 d-\D)\Gamma(d-\D+l-1)} \,,\,\,\,\,I_{\m\n}(x)
\,\,= \,\,  \d_{\m\n}-2\frac{x_{\m}x_{\n}}{x^2}\,,
\ea
the amputated vertex function appearing in (\ref{beta}) is obtained
from (\ref{mvertex}) by the replacements
\ba
\Delta &\rightarrow & d-\Delta\,,\\
\xi_{\mu}(1,3;\5) & \rightarrow & \xi_{\mu}(2,4;\5)\,.
\ea
Then from (\ref{beta}) we obtain
\bea
\label{beta1}
\beta_{\tD}(x_1,x_3;x_2,x_4;\D,l) & =
&\beta_{\tD;\D,l}\frac{1}{(x_{13}^2)^{\tD-\frac12\D}
  (x_{24}^2)^{\tD-\frac12 
    d+\frac12\D}} \nonumber \\
& & \times{}\int\rmd^d
x_{\5}\frac{\Bigl\{e_{\m_1}\cdots e_{\m_l}-{\rm
    traces}\Bigl\}\Bigl\{e'_{\m_1}\cdots e'_{\m_l}-{\rm
    traces}\Bigl\}}{(x_{1\5}^2
  x_{3\5}^2)^{\frac12\D}(x_{2\5}^2x_{4\5}^2)^{\frac12 
    d-\frac12\D}}\,,\\
\beta_{\tD;\D,l} &= & \frac{g^2_{\tD\tD\D,l}}{C_{\D,l}}
\frac{2^{2\tD+\frac12 d+\frac12 
    l}\Gamma(\tD-\frac12\D +\frac12 l)\Gamma(\tD+\frac12\D+\frac12
  l-\frac12 d)}{(2\pi)^d \Gamma(\frac12 d-\tD+\frac12\D+\frac12
  l)\Gamma(d-\tD-\frac12\D+\frac12 l)}\,,\\
e_{\m}&
=&\frac{\xi_{\m}(1,3;\5)}{|\xi^2(1,3;\5)|^{\frac12}}\,\,,\,\,e'_{\m}\,\,=\,\,
\frac{\xi_{\m}(2,4;\5)}{|\xi^2(2,4;\5)|^{\frac12}}\,\,,\,\,e\cdot
e\,\,=\,\,e'\cdot 
e'\,\,=\,\,1\,.
\ea
The product of the unit vectors $e$ and $e'$ in (\ref{beta}) can be
evaluated in terms of Gegenbauer polynomials $C_l^{\frac12 d-1}(x)$ 
\cite{grad} as
\beq
\label{gegen0}
\Bigl\{e_{\m_1}\cdots e_{\m_l}-{\rm
    traces}\Bigl\}\Bigl\{e'_{\m_1}\cdots e'_{\m_l}-{\rm
    traces}\Bigl\}= \frac{1}{c_l^{(l)}}C_l^{\frac12 d-1}(t)\,,
\eeq
where \cite{ruehl2} 
\ba
\label{gegen}
C_l^{\frac{1}{2}d -1}(t) & = & \sum_{M=0}^{l}c_{l}^{(M)}\,t^M \,,\\
t & = &
\frac{\xi_{\mu}(1,3;5)\xi_{\mu}(2,4;5)}{|\xi(1,3;5)||\xi(2,4;5)|}\,.
\ea
This is derived by observing that the following generating function
for Gegenbauer polynomials
\beq
\label{genen1}
\left(1-2a(\xi\cdot\eta)+a^2\xi^2\eta^2\right)^{1-\frac12
d}\,\,\,\,\,\,\,\,\,\,\xi\,,\eta\in {\cal\bf R}^d\,,
\eeq
is harmonic with respect to both Laplacians $\D_{\xi}$ and
$\D_{\eta}$. The latter property is equivalent to the tracelessness of
the tensors in (\ref{gegen0}).
Note that in (\ref{gegen}) $c_M^{(l)}=0$ if $l-M$ is odd. The argument
$t$ of the Gegenbauer polynomials in (\ref{gegen}) can be expanded in
powers of squares $x_{ij}^2$ as
\beq
\label{texp}
t^M = \left[\frac{x_{1\5}^2 x_{2\5}^2 x_{3\5}^2
    x_{4\5}^2}{4 x_{13}^2x_{24}^2}\right]^{\frac{1}{2}M}
\sum_{n_{i,i+1}\in {\bf N}_0}(-1)^{n_{12}+n_{34}} 
\left(\begin{array}{c} M \\ n_{12} \, n_{23} \, n_{34} \,
    n_{41}\end{array}
\right) \prod_{i=1}^{4}\left(\frac{x_{i,i+1}^2}{x_{i\5}^2
    x_{i+1,\5}^2}\right)^{n_{i,i+1}}\,,
\eeq
with $n_{41}=n_{45}$ and $x_{5\5}= x_{1\5}$. Then, the integration of
(\ref{beta}) can be 
reduced to a finite sum of four-star functions which can be evaluated to
\ba
\label{beta11}
\beta_{\tD}(x_1,x_3,x_2,x_4;\Delta,l) & = &
\beta_{\tD;\D,l}\frac{u^{-\tD}}{(x_{12}^2x_{34}^2)^{\tD}}
\sum_{n,m=0}^{\infty} \frac{u^n(1-v)^m}{n!m!}\nonumber \\
& &\hspace{-4.5cm}\times
\Biggl\{\sum_{M=0}^{l}\frac{c_l^{(M)}}{2^M c_{l}^{(l)}}
  \sum_{n_{i,i+1}\in {\bf N}_0}(-1)^{n_{12}+n_{34}}  
\left(\begin{array}{c} M \\ n_{12} \, n_{23} \, n_{34} \,
    n_{41}\end{array}\right) \,v^{n_{23}}\nonumber \\
& & \hspace{-4.2cm}\times\left[u^{\frac12(\D-M)}\a(\d_2)
\a(\d_4)\a(\D)\frac{(\d_1)_n(\frac12 
    d-\d_2)_n(\d_3)_{n+m}(\frac12
    d-\d_4)_{n+m}}{(\D)_{2n+m}(\tD-\frac12 d+1)_{n}}+{\rm shadow
    \,\,term} \right] \Biggl\}\,.
\ea
where 
\bea
i\in(1,3) & & \d_i=\sfrac12(\D-M)+n_{i-1,i}+n_{i,i+1}\,,\\
i\in(2,4) & & \d_i=\sfrac12 d-\sfrac12(\D+M)+n_{i-1,i}+n_{i,i+1}\,.\\
\ea
Performing the final summations we can bring the result into the form
\ba
\label{betafin}
\beta_{\tilde{\Delta}}(x_1,x_3,x_2,x_4;\Delta,l) &= & \beta_{\tD;\D,l}
\frac{u^{-\tD}}{(x_{12}^2
  x_{34}^2)^{\tilde{\Delta}}}
  \sum_{m,n=0}^{\infty}\frac{u^n(1-v)^m}{n!m!}\Bigl[
    u^{\frac{1}{2}(\Delta-l)}D_{nm}(\Delta,l)+\nonumber \\
& & \hspace {4cm}+ u^{\frac12 (d-\Delta
    -l)}D_{nm}(d-\Delta,l) \Bigl]\,.
\ea
The coefficients $D_{nm}$ are regular functions of $\tD$, $\D$ and
$d$, however their form is quite complicated and we do not present
it here \cite{ruehl2}. The first term on the r.h.s. of (\ref{betafin})
is the full 
contribution of the symmetric traceless tensor field with dimension
$\D$ and rank $l$. This is the relevant term which we need in the
direct channel OPE. The second term corresponds again to the shadow
symmetric singularities and it is in general absent in the AdS
calculations. 

\section{The structure of the OPE in the boundary CFT}

We are now in a position to connect the AdS results of Section 3 with
the CPWA of Section 4. From the results of
Section 3 we conclude that an AdS
four-point amplitude ${\cal G}$ (i.e. an appropriately chosen set of AdS
graphs), has the general form
\ba
\label{adsdec}
{\cal G}(x_1,x_2,x_3,x_4)&=&\frac{1}{(x_{12}^2x_{34}^2)^{\tD}}
\sum_{n,m=0}^{ \infty} \frac{u^n(1-v)^m}{n!m!}\left[
  u^{\frac12\D-\tD}c_{nm} -{\cal A}_{nm}\ln u +{\cal B}_{nm}\right]\,,\\
{\cal A}_{nm} & = & (a^{(1)}_{nm} + \tilde{a}_{nm})\gamma_*^2 \,\,\sim
\,\,O(\gamma_*^2)\,,\\ 
{\cal B}_{nm} & = & C^2_{\tD}(\delta_{m0}+(\tD)_{m})\delta_{0n}
+(b^{(1)}_{nm}+\tilde{b}_{nm})\gamma_*^2\,\,\sim\,\,O(1) \,. 
\ea
The first term on the r.h.s. of (\ref{adsdec}) is the full
contribution of the scalar field $\sigma(x)$ with dimension $\D$. This
is the corresponding field of the scalar $\sigma(\hat{x})$ which
appears in the AdS cubic vertex on (\ref{action}). As a check we can
calculate from (\ref{cftex})
and (\ref{b1c}) the coupling $g_{\tD\tD\D}$ and we find it in
agreement 
with the AdS result (\ref{3ptf}). 
We conclude that this is a generic feature
of the OPE in the boundary CFT obtained form AdS/CFT correspondence;
the CFT scalar fields corresponding to the AdS scalars involved
in the triple bulk vertices, appear intact in the operator
algebra of the boundary CFT. \footnote{Note that a cubic vertex like the one in
  (\ref{action}) can be used to calculate the four-point function of
  the scalar field $\sigma(x)$ in the boundary CFT. In this case,
  some connected contributions will come from ``box-graphs'' which are
  $O(\gamma_*^4)$. In principle, such graphs can be calculated using
  the techniques of the present work, as was done in CFT models
  in in $2<d<4$ \cite{ruehl1,ruehl2,tassos1}.} We expect that
this feature holds true for general tensor fields which are involved
in AdS cubic vertices, however the general proof is still
missing. Such a structure of the boundary OPE is robust
against the inclusion of quartic couplings, since it is trivial to
show that general AdS
``star-graphs'' give contributions 
of the form (\ref{b2b3}). Our
results are consistent with the anticipated non-renormalization
\cite{seiberg,freedman4,tassos3} of all operators in the
boundary CFT which correspond to 
the Kaluza-Klein modes of the bulk supergravity theory.

Next we turn to the remaining two terms on the r.h.s. of
(\ref{adsdec}). Following earlier works of one of the authors
\cite{ruehl1,ruehl2}, these terms can be matched to a
conformally invariant OPE in the direct channel which includes
contributions from infinite towers of symmetric, traceless tensor
fields with dimensions (we replace hereafter $\D\rightarrow \D_{l,t}$
and $\beta_{\tD;\D,l}\rightarrow \beta_{l,t}$)
\beq
\label{tdims}
\D_{l,t} = 2\tD+l+2t+\eta_{l,t}\,,
\eeq
where $l$ is the tensor rank and $t$ is an additional quantum number
called the ``twist''. From Bose symmetry $l\in 2{\bf N}_0$ and then
$t\in {\bf N}_0$. These fields have an ``anomalous'' dimension
$\eta_{l,t}\sim O(\gamma_*^2)$. Inserting (\ref{tdims}) into
(\ref{betafin}) and expanding up to $O(\gamma_*^2)$ we get by
comparing the coefficients of $u^n(1-v)^m$ and $u^n(1-v)^m\ln u$
\ba
\label{final1}
\sum_{l,t}\frac{n!}{(n-t)!}D_{nm}(l,t)\,\beta_{l,t} & = & {\cal B}_{nm}\,, \\
\label{final2}
\sum_{l,t}\frac{n!}{(n-t)!}D_{nm}(l,t)\,\sfrac12\eta_{l,t}\,\beta_{l,t}
& = &
-{\cal A}_{nm}\,.
\ea
To proceed we introduce the lexicographic order in the sequences of
labels $(l,t)$ and $(m,n)$ as follows (say for $(m,n)$): we define
\beq
\label{lex1}
(m_1,n_1) < (m_2,n_2)\,,
\eeq
if either
\beq
\label{lex2}
m_1+2n_1<m_2+2n_2\,,
\eeq
or 
\beq
\label{lex3}
m_1+2n_1= m_2+2n_2 \,\,\,\,\,{\rm with}\,\,\,\,\,n_1<n_2\,.
\eeq
Moreover, we consider only those equations in (\ref{final1}),
(\ref{final2}) with
\beq
m+2n\in 2 {\bf N}_0\,.
\eeq
The remaining equations are then interpreted as constraints which have
to be satisfied 
to ensure consistency of our scheme. This has been checked to high
orders in the case of the $O(N)$ vector model in $2<d<4$
\cite{ruehl1,ruehl2}, and in principle can also be done in the
explicit in the case of AdS$_{5}$/CFT$_4$ correspondence using e.g. the
results of \cite{gleb2} \footnote{Work in progress}. Then we note that
the relevant labels $(m,n)$ and $(l,t)$ appear in an ordered sequence
as
\beq
\label{ordered}
(0,0)\,,(2,0)\,,(0,1)\,,(4,0)\,,(2,1)\,,(0,2)\,,(6,0)\,,....\,.
\eeq
This sequence can be mapped into the natural numbers maintaining the
order. Using then 
\ba
\label{matrix1}
\tilde{D}_{(m,n),(l,t)} & = &\frac{n!}{(n-t)!}D_{nm}(l,t)\,,\\
\tilde{{\cal B}}_{(m,n)} & = & {\cal B}_{nm}
\,\,\,\,\,,\,\,\,\,\tilde{{\cal A}}_{(m,n)} \,\,=\,\, -2{\cal A}_{nm}\,,
\ea
we can write (\ref{final1}), (\ref{final2}) in a matrix notation as
\ba
\label{matrix2}
\sum_{s}\tilde{D}_{rs}\,\beta_{s} & = & \tilde{{\cal B}}_{r}\,, \\
\label{matrix3}
\sum_{s}\tilde{D}_{rs}\,\beta_{s}\,\eta_s & = & \tilde{{\cal A}}_{r}\,,
\ea
where $s$ and $r$ now denote the above pairs of indices. These
equations can be solved if we notice that
\beq
\label{triang}
\tilde{D}_{rs} = 0 \,\,\,\,\,\,{\rm if}\,\,\,\,\,r<s\,,
\eeq
i.e. $\tilde{D}_{rs}$ is a triangular matrix. Moreover we have
\beq
\tilde{D}_{rr}\neq 0\,,
\eeq
and this allows to write the general solution of (\ref{matrix1}) and
(\ref{matrix2}) as
\bea
\label{ff1}
\beta_{s} & = & \sum_{1\leq r \leq
  s}(\tilde{D}^{-1})_{sr}\tilde{B}_r\,,\\
\label{ff2}
\beta_{s}\,\eta_{s} & = & \sum_{1\leq r \leq
  s}(\tilde{D}^{-1})_{sr}\tilde{A}_r\,.
\ea
The above constitute, in principle,  the general solution to the
problem of evaluating 
the couplings and anomalous dimensions of all scalar and tensor fields
which appear in the OPE of the boundary CFT. These equations have been
shown to work in the case of the conformally invariant $O(N)$ vector
in $2<d<4$ \cite{ruehl2}. Given the values of the dimensions $\tD$ and
$\D$, they can also be applied to any form of AdS/CFT
correspondence. However, some technical complications might arise in
explicit calculations due to our choice of the normalization of
the two- and three-point functions of tensor fields (\ref{t2ptf}) and
(\ref{mvertex}) (see for example \cite{herzog}). For specific values
of the tensor dimensions and rank, 
it may seem that the normalization constants and coupling include
divergences or zeroes. This is related e.g. to similar problems encountered
in calculations of scalar extremal correlators
\cite{freedman5}. Nevertheless, it can be easily shown \cite{ruehl2}
that in all cases the final formulae (\ref{ff1}) and (\ref{ff2}) yield
  regular results once an overall normalization is suitably chosen.

\section{Summary and Outlook}

In this work we studied a conformally invariant scalar
four-point function which is obtained from AdS$_{d+1}$/CFT$_d$
correspondence. We 
calculated the AdS scalar exchange graphs both in the direct and the
crossed channels and demonstrated that they give analytic results. This
is necessary in order that the corresponding four-point function in
the boundary CFT admits an OPE. We then presented a general procedure to
obtain the contribution of scalar and tensor fields in conformally
invariant four-point functions. Interpreting the logarithms coming
from the AdS calculation as anomalous dimensions of scalar and tensor
fields in the boundary CFT, we were then able to present a general
method for their evaluation by matching the AdS results with the
conformal OPE. As a by-product, we presented some highly
non-trivial formulae for the analytic continuation of generalized
$_3F_2$ hypergeometric functions.  

Our results indicate a possible general form for the OPE of the
boundary CFT. Namely, it seems that AdS scalar fields which are
involved in cubic vertices in the bulk, manifest themselves intact in
the operator 
algebra of the boundary CFT. This essentially means that once their 
dimensions and couplings have been fixed from the supergravity
reduction, they will not change at any order in the ``Witten graph''
expansion. The latter gives, according to the Maldacena conjecture
\cite{maldacena}, the strong coupling limit of the boundary
CFT. Therefore, if these fields appear also in a free-field realization  of
the boundary CFT, we conclude that they are non-renormalized. This is
in agreement with the general view \cite{seiberg,freedman4,tassos3}
that the Kaluza-Klein modes of the compactified supergravity, which
are actually the ones appearing in the cubic AdS vertices, are not
renormalized. Fields which appear {\it either} in the strong {\it or}
in the free-field realization of the boundary CFT may respectively be
{\it either} string modes {\it or} non-chiral primaries, the latter in
the case of ${\cal N}=4$ SYM$_4$. In this latter case, the scalar and
tensor fields whose 
anomalous dimensions are related to the logarithms of the AdS
calculation correspond to composite ``multi-trace'' operators of
the boundary CFT. Thus is seems that the non-trivial dynamics
connecting the 
strongly and the weakly coupled realizations of ${\cal N}=4$ SYM$_4$ 
is essentially related to the gauge group $SU(N)$.  

Our calculations were performed for general $d$ and general dimensions
of the scalar and tensor fields. We expect that significant
simplifications will occur in specific cases such as e.g. the
AdS$_5$/CFT$_4$ correspondence leading to the ${\cal N}=4$ $SU(N)$
SYM$_4$. However, a technical problem related to a consistent overall
normalization of scalars and tensor fields may still occur in explicit
calculations \cite{herzog}. 

There a several possibilities to use and extend our results. In
the case of type IIB compactifications on AdS$_5\times$S$^4$,
the full bulk action has been recently calculated up to quartic
interaction terms \cite{gleb2}. From that one can possibly calculate
the four-point functions 
of various chiral primary operators in ${\cal N}=4$ SYM$_4$ and study
their OPEs. This would shed new light into the strong coupling dynamics
of the latter theory as one could calculate the anomalous dimensions
of non-protected multi-trace operators. From such calculations one
could also deduce interesting results for the energy-momentum tensor
or other conserved currents of the theory. Another possible
application of our result would be in 
the case of AdS$_4$/CFT$_3$ correspondence
\cite{gualtieri}. The calculation of anomalous dimensions in this case
is of particular interest, as it may be compared with existing results
for the anomalous dimensions of various operators in three-dimensional CFTs
\cite{ruehl1,ruehl2,tassos1}. 
 
One additional point which needs further investigation is that the expansion
(\ref{adsdec}) may not be the whole story in a 
perturbatively defined theory in the following sense. In general,
there exists various tensors $(\Delta_1,l_1)$, $(\Delta_2,l_2)$ with
\ba
l_1 & = &l_2\\
\Delta_1 & = &\Delta_2 + {\rm perturbative \,correction \,terms}
\ea
The resolution of such ``almost-degeneracy''requires a sufficiently
high perturbative order and the description of $n$-point functions
with $n>4$. For conformal sigma models in $1/N$ expansion a method of
resolution has been developed in \cite{ruehl2}. By use of
combinatorics it is possible to determine the ``maximal degrees'' of
``almost degeneracy'' using the argument that as $N\rightarrow \infty$
we obtain a free-field theory. These maximal degrees can 
be really trusted only for small $l$ \cite{ruehl2}. Before the
question of 
``almost degeneracy'' is resolved, an expansion such as (\ref{adsdec}) can
only give {\it averaged} anomalous dimensions.

\section*{Acknowledgements}

The work of A.C.P. is supported by an Alexander-von-Humboldt
Research Fellowship.

\appendix{AdS exchange graphs in the direct channel}

The starting point of our calculations is formula (\ref{mb1}). This
can be obtained from (\ref{bgraphs}) either by using Symanzik's
technique \cite{symanzik,ruehl1,tassos1} or following \cite{liu}. To
evaluate the Mellin-Barnes integrals we find it convenient to use the
non-terminating form of Saalschutz's theorem (Eq. 4.3.4.2 of
\cite{slater}) for the Saalschutzian generalized hypergeometric
function $_3F_2$ to obtain
\ba
\label{ap1}
B(x_1,x_3;x_2,x_4) & = & \frac{\kappa\tilde{\kappa}}{( x_{12}^2
  x_{34}^2)^{\tD}} \int_{{\cal C}}\Biggl[ \frac{\rmd s}{2\pi{\rm i}} \frac{
  \Gamma(-s)\Gamma(\frac12\D-\tD-s)\Gamma^4(\tD+
  s)\Gamma(s+1)}{\Gamma(2\tD+2s) \Gamma(\frac1 2\D+\tD-\frac12
  d+1+s)}\nonumber \\
& & \hspace{1cm} \times\,u^s\, {}_2F_1(\tD+s,\tD+s;2\tD+2s;1-v)\Biggl]
\nonumber \\   
 & & \hspace{-1cm} -
\frac{\kappa}{( x_{12}^2 
  x_{34}^2)^{\tD}}\int_{{\cal C}}\frac{\rmd s}{2\pi{\rm
    i}}\Biggl[\Gamma^2(-s)
\frac{\Gamma^4(\tD+s)}{\Gamma(2\tD+2s)} \frac{1}{(s+1)(\frac12\D+\tD-\frac12
  d)} \nonumber \\
& & \times\,{}_3F_2(\sfrac12\D-\tD+1,
\sfrac12\D+\tD-\sfrac12 d+1+s,1;s+2,\sfrac12\D+\tD-\sfrac12
d+1;1)\nonumber\\
& &\hspace{1cm}\times\, u^s\,{}_2F_1(\tD+s,\tD+s;2\tD+2s;1-v)\Biggl]\,,
\ea
where
\beq
\tilde{\kappa} = \frac{\Gamma^2(\frac12\D+\tD-\frac12
  d)\Gamma(\tD-\frac12\D)}{ 
  \Gamma(\frac12\D-\tD+1)\Gamma(2\tD-\frac12
  d)}\,.
\eeq
The contour in the first term on the r.h.s of (\ref{ap1}) encloses only
single poles coming from $\Gamma(-s)$ and $\Gamma(\frac{1}{2}\D-\tD-s)$ at
the points $s=n$ and $s=\frac12\D-\tD+n$, $n=0,1,2,...$
respectively. After expanding the hypergeometric function $_2F_1$, the
result obtained form the first term on the r.h.s. of (\ref{ap1}) is a
double series in the variables $u$ and $(1-v)$. 

The second term on the r.h.s. of (\ref{ap1}) is of the general form
\beq
\label{ap2}
f(u)=\int_{{\cal C}}\frac{\rmd s}{2\pi{\rm i}}\Gamma^2(-s)\,g(s)\,u^s\,,
\eeq
where the function $g(s)$ does not have poles in the right half
plane.  To evaluate the integral we can choose a
regularization method in order to disentangle the {\it two} infinite
series of coincident poles coming from $\Gamma^2(-s)$. For example, we
can choose to shift {\it one } of the infinite series of poles by an
infinitesimal parameter $\e$ which will be set to zero {\it after} the
evaluation of the Mellin-Barnes integral. Setting then
$\Gamma^2(-s) \rightarrow  \Gamma(-s)\Gamma(-s+\epsilon)$ in (\ref{ap1}) we
obtain
\ba
\label{ap3}
f(u) & = & \frac{\partial}{\partial\e}\left[
  \sum_{n=0}^{\infty}\frac{u^{n-\e}
    \,g(n-\e)}{\Gamma^2(n+1-\e)}\right]_{\e=0} \nonumber \\ 
&= & \sum_{n=0}^{\infty}\frac{u^n}{(n!)^2}\left[
  2\psi(n+1)g(n)-g(n)\,\ln u -\frac{\rmd}{\rmd\xi}[g(\xi)]_{\xi=n}\right]\,.
\ea
Using then (\ref{ap3}) we can evaluate the second term on the
r.h.s. of (\ref{ap1}). The result of the lengthy calculation is given
in (\ref{b1a})-(\ref{b1c}) in the text. It is amusing that we are
able to write the final formulas for the coefficients $a_{nm}$,
$b_{nm}$ and $c_{nm}$ in terms of terminating series, by virtue of
the following transformation
\bea
\label{ap4}
&&\hspace{-2cm}_3F_2(\sfrac12\D-\tD+1,\sfrac12\D+\tD-\sfrac12
d+1+n,1;n+2,\sfrac12\D+\tD-\sfrac12 d+1;1) =\nonumber \\
&&\hspace{1cm}
=\frac{\Gamma(n+2)\Gamma( \frac12\D+\tD-\frac12 d+1)}{ (\tD-\frac12\D)
\Gamma(\frac12\D+\tD-\frac12 d+1+n)}\nonumber \\
& &\hspace{1.5cm}\times{}_3F_2(1-\sfrac12\D-\tD+\sfrac12
d,\tD-\sfrac12\D, -n;\tD-\sfrac12\D+1,1;1)\,,
\ea
which is a consequence of the two-term relation Eq.(2.3.3.7)  of
\cite{slater}.

\section{AdS exchange graphs in the
  crossed channel} 

Here we present the essential steps for the crossed channel
calculation corresponding to the interchange
$u\leftrightarrow v$ in (\ref{ap1}). We first observe that the
argument of the $_2F_1$ hypergeometric functions becomes $(1-u)$ and
in order to transform it into a series in the variable $u$ we need to
use a Kummer transformation e.g. Eq. 9.131.2 of \cite{grad}. Due to
the form of the $_2F_1$ function involved in the transformation, we
encounter poles in the r.h.s. of Eq. 9.131.2 of
\cite{grad} i.e. this is a {\it degenerate} transformation. This is a
consequence of the analytic continuation 
implied by the above Kummer transformations. To obtain the result we
may use, for example, the Mellin-Barnes representation for the $_2F_1$
function and appropriately  regularize the coincident poles as in
(\ref{ap3}) above. It is nevertheless simple to write
\ba
\label{ap5}
_2F_1(\tD+s,\tD+s;2\tD+2s;1-u) &\equiv& \left.
  {}_2F_1(\tD+s,\tD+s;2\tD+2s+\e;1-u)\right|_{\e\rightarrow 0}
\nonumber \\
&  & \hspace{-2cm}=\Biggl[\frac{\Gamma(2\tD+2s+\e)\Gamma(\e)}{
  \Gamma^2(\tD+s+\e)}{}_2F_1( 
\tD+s,\tD+s;1-\e;u)\nonumber \\
& & \hspace{-2cm} +u^{\e}\,\frac{\Gamma(2\tD+2s+\e)\Gamma(-\e)}{
    \Gamma^2(\tD+s)} 
{}_2F_1( \tD+s+\e,\tD+s+\e;1+\e;u)\Biggl]_{\e\rightarrow 0}\nonumber \\
&  & \hspace{-4cm}= \frac{\Gamma(2\tD+2s)}{\Gamma^4(\tD+s)} \sum_{n=0}^{\infty}
\frac{u^n}{(n!)^2}\Gamma^2(\tD+s+n)\left[ -\ln
  u+2\psi(n+1)-2\psi(\tD+s+n)\right] \nonumber \\
 &  & \hspace{-2cm}= \frac{\Gamma(2\tD+2s)}{\Gamma^4(\tD+s)}
 \sum_{n=0}^{\infty} 
\frac{u^n}{(n!)^2}\,\left.
  {\cal
  D}_n(\frac{\partial}{\partial\xi})\,\Gamma^2(\tD+s+n+\xi)\right|_{\xi=0} \,,
\ea
where 
\beq
{\cal D}_m(\frac{\partial}{\partial\xi})  = -\ln u +2\psi(m+1)
  -\sfrac{\partial}{\partial \xi}\,.
\eeq
Using then (\ref{ap5}) we obtain from (\ref{ap1}) 
\ba
\label{ap6}
B(x_1,x_4;x_2,x_3) & = &
\frac{\kappa\hat{\kappa}}{(x_{12}^2x_{34}^2)^{\tD}} \Biggl[  
  \sum_{m=0}^{\infty} \frac{u^m}{(m!)^2}{\cal
    D}_m(\frac{\partial}{\partial\xi}) \Bigl( v^{\frac12 \D-\tD}
  f_1(v,\xi) -f_2(v,\xi) -f_3(v,\xi)\Bigl)\Biggl]_{\xi=0}\\
f_1(v,\xi) &=& \frac{\Gamma^2(\frac12 \D+m+\xi)}{\Gamma(
  \D-\frac12 d+1)} 
  {}_2F_1(\sfrac12 \D+m+\xi,\sfrac12 \D+m+\xi; \D-\sfrac12 d+1;v
  ) \label{f1}\,,\\
f_2(v,\xi) & = & \frac{\Gamma^2(\tD+m+\xi)}{\Gamma(\frac12
  \D+\tD-\frac12 d+1)\Gamma(\tD-\frac12 \D+1)}\nonumber \\
 & & \times {}_3F_2\left( \tD+m+\xi,
  \tD+m+\xi, 1;\sfrac12 \D+\tD-\sfrac12 d+1,\tD-\sfrac12
  \D+1;v\right),\label{f2}\\
f_3(v,\xi) &= & \frac{\Gamma(2\tD-\frac12 d)}{\Gamma^2( \tD-\frac12\D)
  \Gamma^2(\tD-\frac12 \D-\frac12 d)} \int_{{\cal C}}\frac{\rmd s}{2\pi{\rm i}}
\frac{\Gamma^2(-s)\Gamma^2(\tD+m+s+\xi)}{ (s+1)(\tD+\frac12 \D-\frac12
  d)}v^s  \nonumber \\ 
& & \hspace{-1cm} \times{}_3F_2\left(\sfrac12 \D-\tD+1,\sfrac12
  \D+\tD-\sfrac12 
  d+1+s,1; s+2,\sfrac12 \D+\tD-\sfrac12 d+1;1\right), \label{f3}
\ea
with
\beq
\hat{\kappa} = \frac{\Gamma^2(\frac12\D+\tD-\frac12 d)\Gamma^2(
  \tD-\frac12\D)}{\Gamma(2\tD-\frac12 d)}\,.
\eeq
First we show that
$f_3(v,\xi)$ does 
not contain non-analytic terms as
$v\rightarrow 1$. Using (\ref{ap3}) we can write 
\beq
\label{f3epsilon}
f_{3}(v,\xi) =
\frac{\partial}{\partial\e}\left[\sum_{n=0}^{\infty}\frac{v^{n-\e}
    g(n-\e,\xi)}{\Gamma^2(n+1-\e)} \right]_{\e=0}\,.
\eeq
The possible
non-analytic terms as $v\rightarrow 1$ in 
(\ref{f3epsilon})  are determined by the large-$n$
asymptotics of the ratio $g(n-\e,\xi)/\Gamma^2(n+1-\e)$. This in turn
can be found using the Stirling formula below \cite{grad}
\ba
\label{stirl}
\left.\frac{\Gamma(a+r+1)}{\Gamma(r+1)}\right|_{r\rightarrow \infty}
&\approx& \exp\left[a\ln(r+1)
    +\sum_{k=1}^{\infty}\frac{{\cal P}_{k+1}(a)}{(r+1)^k}\right]\,,
\\
{\cal P}_{k+1}(a) & = & \nonumber \frac{(-1)^{k+1}}{k}\sum_{l=0}^n
\left(\begin{array}{c} n \\ l\end{array}\right)
\frac{B_l}{n-l+1}(a-1)^{n-l+1}\,\,\,,\,\,\,a\in{\cal\bf C}\\ 
& = &
\frac{(-1)^{k+1}}{k}\sum_{l=0}^{a-1}{l^k}\,\,\,,\,\,\,a\in{\cal\bf N}\,. 
\label{Pi}
\ea
where $B_l$ are the Bernoulli numbers (Secs. 9.61, 9.71 of \cite{grad}). This
enables us to obtain an asymptotic expansion for $_3F_2$ in 
(\ref{f3}) by virtue of 
Eq. 4.3.4.2 of \cite{slater} setting (in Slater's notation)
$c=\frac12 \D+\tD-\frac12 d+1+s$, $e=s+2$. In this way we obtain
\beq
\label{goverG}
\left. \frac{g(n-\e,\xi)}{\Gamma^2(n+1-\e)}\right|_{n\rightarrow
  \infty} \approx \sum_{i=1}^2 \sum_{\lambda}^{\infty} 
\tilde{\sigma}_{\lambda ,i}
\frac{\Gamma(A_i+n+1-\e-\lambda)}{ \Gamma(n+1-\e)}\,,
\eeq
for some parameters $A_i$, which depend among others on $m$ and
$\xi$. The coefficients $\tilde{\sigma}_{\lambda,i}$ can in principle be
explicitly determined in terms of the Bernoulli numbers \cite{HPR1},
but in this case we only require their existence. Then, by
virtue of (\ref{goverG}) we obtain the non-analytic terms
\ba
\label{f3final}
 \left. \sum_{n=0}^{\infty}\frac{v^{n-\e}
    g(n-\e,\xi)}{\Gamma^2(n+1-\e)}\right|_{n.a.} &
\approx & \left. \sum_{i=1}^2\sum_{\lambda}^{\infty} \tilde{\sigma}_{\lambda,i}
\,v^{-\e}\frac{\Gamma(A_i
  +1-\e-\lambda)}{\Gamma(1-\e)}{}_2F_1(A_i+1-\e-\lambda,1;1-\e;v)
\right|_{n.a.} 
\nonumber \\
& \approx & \sum_{i=0}^2\sum_{\lambda}^{\infty} \tilde{\sigma}_{\lambda,i}
\,\Gamma(A_i+1-\lambda) (1-v)^{-A_i-1+\lambda} \,,
\ea  
where to get the second line of (\ref{f3final}) we used a Kummer
transformation and have omitted the analytic terms. The crucial point
is now that the  
non-analytic terms are independent of the parameter $\e$ and
therefore  drop out  
when we substitute (\ref{f3final}) into (\ref{f3epsilon}).

We next consider the function
\beq
\label{ap7}
v^{\frac12\D-\tD}f_1(v,\xi)-f_2(v,\xi)\,.
\eeq
The assertion that (\ref{ap7}) is holomorphic at $v=1$ is a mathematical
theorem that will be proved first. Our presentation of the proof
follows the review article of N\/orlund \cite{norlund}. 

Using a Kummer transformation we can see that the first term in
(\ref{ap7}) involves the non-analytic part
\beq
\label{naf1}
(1-v)^{1-\frac12 d-2m-2\xi}\Gamma(\sfrac12
d-1+2m+2\xi) \sum_{r,n=0}^{\infty}\frac{
  (1-v)^{r+n}}{r!n!}\frac{(\tD-\frac12\D)_r (\frac12 \D-\frac12
    d+1-m-\xi)_{n}^2}{(2-\frac12 
  d-2m-2\xi)_{n}}\,, 
\eeq 
where we have used the binomial expansion for $v^{\frac12\D-\tD}$. 
We will show that the second term in (\ref{ap7}) involves a
non-analytic term which just cancels the above contribution.  

The second term of (\ref{ap7}) can be brought into the form (Eq. 1.13
of \cite{norlund})
\ba
\label{ap8}
\Phi_s(v)&=&
v^{\gamma_s}\frac{\left[\prod_{r=1}^{3}\Gamma(\a_r
+\gamma_s)\right]}{\left[\prod_{ r\neq s}\Gamma( \gamma_r
-\gamma_s+1)\right]} \nonumber \\
& & \times {}_3F_2(\a_1+\gamma_s,\a_2+\gamma_s,\a_3+\gamma_s;
\gamma_s-\gamma_1+1,\gamma_s-\gamma_2+1,\gamma_s-\gamma_3+1;v)^{\dagger}\,,
\ea
where $s=1,2,3$ and the $^{\dagger}$ indicates that the argument
$\gamma_s-\gamma_s+1=1$ is left out. In this representation
(\ref{ap8}) satisfies the differential equation
\beq
\label{ap9}
\left[ {\cal Q}(v\frac{\rmd}{\rmd v})-v{\cal R}(v\frac{\rmd}{\rmd
    v})\right]\,f_2(v)=0\,,
\eeq
where ${\cal Q}(x)$ and ${\cal R}(x)$ are the polynomials
\beq
\label{ap10}
{\cal Q}(x) = \prod_{i=1}^{3}(x-\gamma_i) \,\,\,\,,\,\,\,\, {\cal
  R}(x) = \prod_{i=1}^{3}(x-\a_i)\,. 
\eeq
The linear differential equation (\ref{ap9}) has a regular
singularity at $v=1$. As shown in \cite{norlund} it possesses there
one singular solution $\chi_3(v)$ and two regular solutions
$\rho_3^{(1)}(v)$ and $\rho_3^{(2)}(v)$ which form a basis (``basis
theorem''). It follows therefore that
\beq
\label{ap11}
f_2(v) = {\cal C}_s\chi_3(v)
+\sum_{i=1}^2{\cal}{\cal C}_r^{(i)}\rho_3^{(i)}(v)\,\,\,,\,\,\,0<v<1\,.
\eeq
The coefficients ${\cal C}_s$, ${\cal C}_r^{(1)}$ and ${\cal
C}_r^{(2)}$ will be determined
later. (\ref{ap11}) constitutes a 
Kummer transformation for the generalized hypergeometric function
$_3F_2$. 

The identification of the parameters in (\ref{ap8}) is not unique. A
convenient choice for our purposes is to set
\ba
\label{ap12}
s=3\,, & &\a_1=\a_2=\tD+m+\xi\,,\,\,\,\,\,\,\,\,\a_3=1\,, \nonumber \\
\gamma_1 = \sfrac12 d-\sfrac12\D-\tD\,, & &\gamma_2=\sfrac12\D-\tD\,,
\,\,\,\,\,\,\,\,\, \gamma_3=0\,.
\ea
With this choice we obtain from Eq. 1.15 of \cite{norlund}
\ba
\beta_1 & = & \sfrac12\D-\sfrac12 d-m-\xi\,,\\
\label{beta3}
\beta_2 = \beta_3 & = &  1-2m-2\xi-\sfrac12 d\,,
\ea
and from Eq. 1.33 and 2.11 ibid.
\ba
\label{ap13}
\chi_3(v) & = & \sum_{n=0}^{\infty}\frac{{\cal
    C}^{(3)}_{n,3}}{(\beta_3+1)_{ n}}(1-v)^{\beta_3+n}\,,\\
\label{ap14}  
\frac{{\cal
    C}^{(3)}_{n,3}}{(\beta_3+1)_{ n}} & = & \sum_{r=0}^{n}\frac{
  (1+\frac12\D-\frac12 d-m-\xi)_r^2 (\tD-\frac12 \D)_{n-r}}{
  r!(n-r)!(2-2m-2\xi-\frac12 d)_r} \,.
\ea
Therefore, from (\ref{ap11}) we see that the singular part of
$f_2(v,\xi)$ cancels the singular 
part (\ref{naf1}) of $f_1(v,\xi)$ if
\beq
\label{ap15}
{\cal C}_s= \Gamma(2m+2\xi+\sfrac12 d-1)\,.
\eeq
To show this we can use the method of ``large order expansion'' of the
coefficients of $f_2(v,\xi)$ in powers of $v$.  Namely, writing
\beq
\label{naf2}
f_2(v,\xi)= \sum_{n=0}^{\infty} \frac{v^n}{n!}\frac{\Gamma^2(\tD+m+\xi+n)
  \Gamma(n+1)}{ \Gamma(\frac12 \D+\tD-\frac12 d+1+n)\Gamma(
  \tD-\frac12 \D +1+n)}\,,
\eeq
we can use Stirling's formula (\ref{stirl}) to obtain
\beq
\label{stappl}
\left. \frac{\Gamma^2(\tD+m+\xi+n)
  }{ \Gamma(\frac12 \D+\tD-\frac12 d+1+n)\Gamma(
  \tD-\frac12 \D +1+n)}\right|_{n\rightarrow \infty} \approx
\sum_{\lambda=0}^{\infty} \sigma_{\lambda}
\frac{\Gamma(\beta-\lambda+n+1)}{ \Gamma(n+1)}\,,
\eeq
where $\beta$ is defined in (\ref{sigmas1}). From (\ref{stappl}) the non-analytic part of $f_2(v,\xi)$ can be found as
\ba 
\label{naf20}
\left. \sum_{n=0}^{\infty} \frac{v^n}{n!}\frac{\Gamma^2(\tD+m+\xi+n)
  \Gamma(n+1)}{ \Gamma(\frac12 \D+\tD-\frac12 d+1+n)\Gamma(
  \tD-\frac12 \D +1+n)}\right|_{n.a.} & \approx &
\sum_{\lambda}^{\infty} \sigma_{\lambda} 
\left[ \sum_{n=0}^{\infty} \frac{\Gamma(\beta
    -\lambda+n+1)}{\Gamma(n+1)} v^n \right]\nonumber \\
 &  & \hspace{-5cm}\approx\sum_{\lambda=0}^{\infty} \sigma_{\lambda}
 \,\Gamma(\sfrac12 d -1 
 +2m+2\xi-\lambda)(1-v)^{1-\frac12 d-2m-2\xi +\lambda}\,.
\ea 
Since from (\ref{ap14}) and (\ref{stappl}) 
\beq
\sigma_0 = \frac{{\cal C}^{(3)}_{0,3}}{(\beta_3+1)_0}=1\,,
\eeq
then (\ref{ap15}) follows.

In our previous work \cite{HPR1}, being unaware of \cite{norlund} we
went on to show that the coefficients $\sigma_{\lambda}$ in
(\ref{stappl}) equal the r.h.s of (\ref{ap14}) for all $\lambda$. 
The explicit formula for the $\sigma_{\lambda}$ is obtained recursively from 
\ba
\label{sigmas}
& & \hspace{-1cm}\sum_{\lambda=0}^{\infty}\frac{\sigma_{\lambda}}{(n+1)^{\lambda}}
  \exp\left[\sum_{k=1}^{\infty} \frac{{\cal
      P}_{k+1}(\beta-\lambda) }{(n+1)^{k}}\right] \begin{array}{c}
  {}\\ \approx \\\scriptstyle{n\rightarrow\infty}\end{array} 
\exp\left[\sum_{k=1}^{\infty} 
    \frac{2{\cal P}_{k+1}(t_1)-{\cal P}_{k+1}(t_2)
      -{\cal P}_{k+1}(t_3)}{(n+1)^{k}} \right],\\
& & \hspace{-1cm}t_1 =  \tD+m+\xi-1\,,\,\,\,t_2= \tD-\sfrac12
\D\,,\,\,\,t_3= \sfrac12 \D +\tD-\sfrac12 d \,,\\
& & \label{sigmas1} 
\hspace{-1cm}\beta  =  2t_1-t_2-t_3 = \sfrac12 d-2+2m+2\xi\,,
\ea 
by matching the powers of $1/(n+1)$ on both sides of
(\ref{sigmas}).
Then, from (\ref{ap14}) and (\ref{naf20}) we have to prove that
\beq
\label{final}
\sum_{\lambda=0}^{\infty} \sigma_{\lambda}
\frac{(-1)^{\lambda}(1-v)^{\lambda}}{ (2-\frac12 d-2m-2\xi)_{\lambda}} =
\sum_{k,l=0}^{\infty} \frac{(1-v)^{k+l}}{k!l!}\frac{(\tD-\frac12
  \D)_{l}(\frac12 \D-\frac12 d+1-m-\xi)_k^2}{ (2-\frac12 d-2m-2\xi)_k}\,.
\eeq
To prove this we will show that
\beq
\label{A1}
\sigma_{\lambda} = \sum_{k=0}^{\lambda}\frac{(-1)^k}{k!(\lambda -k)}
(t_2)_{\lambda -k} (t_3-t_1)_{k}^{2}(\beta -\lambda +1)_{\lambda -k}\,.
\eeq
The proof is based on the observation that (\ref{stappl}) and
(\ref{sigmas}) have dual forms. For (\ref{stappl}) we obtain
\ba
\label{A2}
\frac{\Gamma^2(t_1+z+1)}{\Gamma(t_2+z+1)\Gamma(t_3+z+1)} &
\begin{array}{c} {} \\ \approx \\ \scriptstyle{z\rightarrow
-\infty}\end{array}
 &\frac{\sin\pi(t_2+z)\sin\pi(t_3+z)\sin\pi(\beta+z)}{
 \sin^2\pi(t_1+z)\sin\pi
 z}\nonumber \\
& & \times\sum_{\lambda=0}^{\infty}\sigma_{\lambda}\frac{\Gamma(\beta
 -\lambda+z+1)}{\Gamma(z+1)}\,, 
\ea
while for (\ref{sigmas}) we have
\ba
\label{A3}
\hspace{-1cm}\sum_{\lambda=0}^{\infty}\frac{
(-1)^{\lambda}\sigma_{\lambda}}{(z+1)^{\lambda}}  
  \exp\left[\sum_{k=1}^{\infty} \frac{{\cal
      P}_{k+1}(\beta-\lambda) }{(z+1)^{k}}\right] &\begin{array}{c}
  {}\\ \approx \\\scriptstyle{z\rightarrow -\infty}\end{array}&
 \frac{ \sin^2\pi(t_1+z)\sin\pi
 z}{  
\sin\pi(t_2+z)\sin\pi(t_3+z)\sin\pi(\beta+z)} \nonumber \\
&&\hspace{-0.5cm}\times\exp\left[\sum_{k=1}^{\infty} 
    \frac{2{\cal P}_{k+1}(t_1)-{\cal P}_{k+1}(t_2)
      -{\cal P}_{k+1}(t_3)}{(z+1)^{k}} \right]\,,
\ea
with the same $\sigma_{\lambda}$. We also need the following property
\beq
\label{A4}
{\cal P}_{k+1}(1-t) = (-1)^{k+1}{\cal P}_{k+1}(t)
\eeq
Now we will show that (\ref{A2}) follows from (\ref{A1}). Multiplying
the r.h.s. of (\ref{A1}) by
\beq
\label{A5}
\frac{\Gamma(\beta-\lambda+z+1)}{\Gamma(z+1)}\,,
\eeq
and summing over $\lambda$ we obtain a convergent double sum for
Re$z\rightarrow -\infty$
\ba
\label{A6}
&&\hspace{-1.5cm}
\sum_{k,l=0}^{\infty}\frac{(-1)^k}{k!l!}(t_2)_{l}(t_3-t_1)_{k}^2(\beta-l-k+1)_l  
\frac{\Gamma(\beta-l-k+z+1)}{\Gamma(z+1)} =\nonumber \\
&&\hspace{2cm}=\frac{\sin\pi
z}{\sin\pi(\beta+z)}
\sum_{k=0}^{\infty}\frac{
\Gamma(-z)}{\Gamma(k-\beta-z)}\frac{(t_3-t_1)^2_k}{k!}
{}_2F_1(k-\beta,t_2;k-\beta-z;1) \nonumber \\
& & \hspace{2cm}=\frac{\sin\pi
z}{\sin\pi(\beta+z)}\frac{\Gamma(-z-t_2)}{\Gamma(
-\beta-z-t_2)}{}_2F_1(t_3-t_1,t_3-t_1;-\beta-z-t_2;1)\nonumber \\
  &&\hspace{2cm}=\frac{ \sin^2\pi(t_1+z)\sin\pi z}{  
\sin\pi(t_2+z)\sin\pi(t_3+z)\sin\pi(\beta+z)}
\frac{\Gamma^2(t_1+z+1)}{\Gamma(t_2+z+1)\Gamma( t_3+z+1)}\,.
\ea
Since all the above steps are invertible, (\ref{final}) and (\ref{A1})
are proven {\it q.e.d.}

Next we turn to the analytic part of $f_2(v,\xi)$ in (\ref{ap11}). We
can then write from Sec. 5.7 of \cite{norlund}
\beq
\label{ap16} 
\rho_3^{(1)}(v) = y_{1,2}(v)\,,\,\,\,\,\,\,\,\,\,\rho_{3}^{(2)}(v)=
y_{1,3}(v)\,, 
\eeq
where e.g.
\beq
\label{ap17}
y_{1,2}(v) = \int_{{\cal C}}\frac{\rmd s}{2\pi{\rm i}}\,v^s\,
\frac{\Gamma(\gamma_1-s)
  \Gamma(\gamma_2-s)\left[\prod_{i=1}^{3}\Gamma(\a_i+s)\right]}{
  \Gamma(s+1-\gamma_3)}\,.
\eeq
The contour ${\cal C}$ separates the increasing from the decreasing sequence of
poles. The integral in (\ref{ap17}) converges for   
\beq
2\pi>{\rm arg} \,v>-2\pi\,,
\eeq
which includes the circle
\beq
|1-v|\leq 1\,.
\eeq
Then, an expansion of $y_{1,2}(v)$ in powers of $(1-v)$ is given by
Eq. 5.35 of \cite{norlund} as
\ba
\label{ap18}
y_{1,2}(v) & = & C_1\,v^{\gamma_2}\,\sum_{n=0}^{\infty}
\frac{(\a_1+\gamma_2)_n
  (\a_2+\gamma_2)_n}{n!(\a_1+\a_2+\gamma_1+\gamma_2)_n} \,(1-v)^{n}
\nonumber \\
& &
\hspace{-1cm} \times\,
{}_3F_2(\a_1+\gamma_1,\a_2+\gamma_1,1-\a_3-\gamma_3;\gamma_1-\gamma_3+1,
\a_1+\a_2+\gamma_1+\gamma_2+n;1)\,,\\
C_1 &= &
\frac{\Gamma(\a_1+\gamma_2)\Gamma(\a_2+\gamma_2)\left[\prod_{i=1}^3\Gamma(
  \a_i+\gamma_1)\right]}{
  \Gamma(\a_1+\a_2+\gamma_1+\gamma_2)\Gamma(\gamma_1-\gamma_3+1)}\,. 
\ea
The corresponding results for $\rho_3^{(2)}$ are obtained from above
by the interchange of indices $2\leftrightarrow 3$. 

Now we return to (\ref{ap11}). Shifting the contour in (\ref{ap17}) to
$+\infty$ we obtain
\ba
\label{ap19}
y_{1,2}(v) &=&
\frac{\pi}{\sin\pi(\gamma_2-\gamma_1)}\left[\Phi_1(v)-\Phi_2(v)\right]
\,,\\
\label{ap20}
y_{1,3}(v) &=&
\frac{\pi}{\sin\pi(\gamma_3-\gamma_1)}\left[\Phi_1(v)-\Phi_3(v)\right]\,.
\ea
The corresponding  Mellin-Barnes representation for $\chi_{3}(v)$
reads (Eq. 2.44 of \cite{norlund})
\beq
\label{ap21}
\chi_3(v)=\Gamma(\beta_3 +1)\int_{\cal C}\frac{\rmd s}{2\pi{\rm i}}
\,v^{-s}\, \prod_{r=1}^3
\frac{\Gamma(s+\gamma_r)}{\Gamma(s-\a_r+1)}\,,
\eeq
with the same $\beta$ as in (\ref{beta3}). Shifting the integration
contour above to $-\infty$ we obtain
\beq
\label{ap22}
\chi_3(v) = \frac{\Gamma(\beta_3+1)}{\pi} \sum_{i=1}^{3}\frac{
\left[\prod_{s=1}^3 \sin\pi(\gamma_i+\a_s)\right]}{ \left[ 
\prod_{s\neq i}\sin\pi(\gamma_i -\gamma_s)\right] } \Phi_i (v)\,.
\eeq
The linear system (\ref{ap19}), (\ref{ap20}) and (\ref{ap22}) can be
inverted. For the evaluation of its determinant we need the identity
\ba
\label{ap23}
& &\hspace{-1cm}\sum_{(i,j,k)\,{\rm in\,cyclic \,order}}^3\sin\pi(\gamma_i
-\gamma_j)\prod_{s=1}^3\sin\pi(\gamma_k +\a_s) =\nonumber \\
& &\hspace{1cm}=-\sin\pi(\gamma_1 -\gamma_2)\sin\pi(\gamma_2
-\gamma_3)\sin\pi(\gamma_3 -\gamma_1) \sin\pi\left(\sum_{i=1}^3(\a_i
+\gamma_i)\right) \,.
\ea
Then, for the coefficients in (\ref{ap11}) we obtain
\ba
\label{ap24}
{\cal C}_r^{(1)} & = & \frac{1}{\pi} \frac{\left[
\prod_{s=1}^3\sin\pi(\gamma_2 +\a_s)\right]}{
\sin\pi\beta_3\,\sin\pi(\gamma_3 -\gamma_2)} \,,\\
   {\cal C}_r^{(2)} & = & \frac{1}{\pi} \frac{\left[
\prod_{s=1}^3\sin\pi(\gamma_3 +\a_s)\right]}{
\sin\pi\beta_3\,\sin\pi(\gamma_2 -\gamma_3)}\,,
\ea
and also (\ref{ap15}) again. The latter is a non-trivial check for our
calculations. 

Next we want to derive a Mellin-Barnes integral for the coefficients 
\beq
\label{ap25}
-\tilde{a}_{nm} \ln u + \tilde{b}_{nm}\,,
\eeq
in (29). Here we restrict the derivation only to the contribution
of $B(x_1,x_4;x_2,x_3)$. Expanding  (87) in powers of $u$ we
have to treat (see (21) with $u \leftrightarrow v$)
\ba
\label{ap26}
&&\hspace{-1cm} \frac{\kappa \kappa''}{n!} \int_C \frac{ds}{2\pi i}
\frac{\Gamma^2(-s) 
\Gamma(\frac{1}{2} \Delta -\tilde{\Delta} -s)}{\Gamma(\frac{1}{2}
\Delta+\tilde{\Delta}-\frac{1}{2}d-s)} v^s \nonumber \\ 
&&\hspace{-0.5cm}\times
_3F_2(\frac{1}{2}\Delta+\tilde{\Delta}-\frac{1}{2}d,
\frac{1}{2}\Delta+\tilde{\Delta}-\frac{1}{2}d,
\frac{1}{2}\Delta-\tilde{\Delta}-s; \Delta-\frac{1}{2}d+1,
\tilde{\Delta}+\frac{1}{2}\Delta-\frac{1}{2}d-s;1)\nonumber \\
&&\hspace{-0.5cm} \times \mathcal{D}_n(\frac{\partial}{\partial\xi})
\Gamma^2(\tilde{\Delta}+s+n+\xi)\arrowvert_{\xi=0}. 
\ea 
By the Stirling formula \cite{grad} we know that for $s=\pm i\sigma,
\sigma \rightarrow \infty$ 
\beq
\label{ap27}
\frac{\Gamma^2(-s)\Gamma(\frac{1}{2}\Delta-\tilde{\Delta}-s)\Gamma^2(\tilde{\Delta}+s+n+\xi)}{\Gamma(\frac{1}{2}\Delta+\tilde{\Delta}-\frac{1}{2}d-s)}
= O(e^{-2\pi\sigma})\,.
\eeq
Moreover the $_3F_2$-function in (\ref{ap26}) behaves in the same limit
as a power of $\sigma$
\beq
\label{ap28}
{\rm const.}\, \times \, \sigma^{{\rm max}[0,
\:2\tilde{\Delta}-\frac{1}{2}d-1]} \,.
\eeq
An appropriate three-term relation for $_3F_2(1)$ (\cite{slater},
eqn.(4.3.4.2) with $a=b=\frac{1}{2}\Delta+\tilde{\Delta}-\frac{1}{2}d
, \\ c=\frac{1}{2}\Delta-\tilde{\Delta}-s ,  d=\Delta-\frac{1}{2}d+1 ,
e=\tilde{\Delta}+\frac{1}{2}\Delta -\frac{1}{2}d-s$) and the Stirling
formula \cite{grad} gives a complete asymptotic series expansion for
this $_3F_2$-function with leading term (\ref{ap28}). 
Inserting then the Taylor expansion
\be
\label{ap29}
v^s=\sum_{m=0}^\infty \frac{(1-v)^m}{m!} (-s)_m\,,
\eeq
into (\ref{ap26}) we obtain as contribution to (\ref{ap25}) the
exponentially convergent Mellin-Barnes integral
\ba
\label{ap30}
\frac{\kappa \kappa''}{n!} \int_C \frac{ds}{2\pi i} \frac{\Gamma^2(-s)
\Gamma(\frac{1}{2} \Delta -\tilde{\Delta} -s)(-s)_m
}{\Gamma(\frac{1}{2} \Delta+\tilde{\Delta}-\frac{1}{2}d-s)} \nonumber
\\ \times _3F_2(\frac{1}{2}\Delta+\tilde{\Delta}-\frac{1}{2}d,
\frac{1}{2}\Delta+\tilde{\Delta}-\frac{1}{2}d,
\frac{1}{2}\Delta-\tilde{\Delta}-s; \Delta-\frac{1}{2}d+1,
\tilde{\Delta}+\frac{1}{2}\Delta-\frac{1}{2}d-s;1) \nonumber\\
\times \mathcal{D}_n(\frac{\partial}{\partial\xi})
\Gamma^2(\tilde{\Delta}+s+n+\xi)\arrowvert_{\xi=0}\,, 
\ea
with
\beq
\label{ap31}
\kappa''=\frac{\Gamma^2(\frac{1}{2}\Delta+\tilde{\Delta}
-\frac{1}{2}d)}{\Gamma(\Delta-\frac{1}{2}d+1)}\,.  
\eeq
The holomorphy of the functions $y_{1,2}(v), y_{1,3}(v)$ at $v=1$ can
equally be derived from the exponential convergence of the
Mellin-Barnes integrals (\ref{ap17}).


\begin{thebibliography}{99}
 \bibitem{maldacena}
J. Maldacena, ``The large N limit of superconformal field 
theories and supergravity'', Adv. Theor. Math. Phys. {\bf 2} (1998) 
231-252, hep-th/9711200; 
S.S. Gubser, I.R. Klebanov and A. M. 
Polyakov, ``Gauge theory correlators from noncritical string 
theory'', Phys. Lett. {B428} (1998) 105, hep-th/9802109; 
E. Witten, 
``Anti-de  Sitter space and holography'', Adv. Theor. Math. 
Phys. {\bf 2} (1998) 253-291, hep-th/9802150.

\bibitem{oz}
O.~Aharony, S.S.~Gubser, J.~Maldacena, H.~Ooguri and Y.~Oz,
``Large N field theories, string theory and gravity'',
hep-th/9905111.


\bibitem{polyakov}
A. M. Polyakov, ``The wall of the cave'', Int. J. Mod. Phys. {\bf A14}
(1999) 645, hep-th/9809057. 


\bibitem{haag}
R. Haag {\it Local quantum field theory}, Springer (1992).


\bibitem{rehren}
K. -H. Rehren, ``Algebraic Holography'', hep-th/9905179.

\bibitem{fradkin}
E. S. Fradkin and M. Ya. Palchik, ``New developments in
$d$-dimensional conformal quantum field theory'', Phys. Rep. {\bf 300}
(1998) 1. 

\bibitem{todorov}
G. Mack and I. T. Todorov, Phys. Rev. {\bf D8} (1973), 1764;\\
V. K. Dobrev, V. B. Petkova, S. G. Petrova and I. T. Todorov,
Phys. Rev. {\bf D13} (1976), 887.

\bibitem{viswa}
W. M\"uck and K. S. Viswanathan, ``Conformal field theory correlators
from classical scalar field theory on AdS$_{d+1}$,  Phys. Rev. {\bf
D58} (1998) 041901, hep-th/9804035; ``Conformal field theory correlators from
classical field theory on Anti-de-Sitter space II: vector and spinor
fields, Phys. Rev {\bf D58} (1998) 106006, hep-th/9805145.

\bibitem{freedman0}
D. Z. Freedman, S. D. Mathur, A. Matusis and L. Rastelli, ``Correlation
functions in the CFT(d)/AdS(d+1) correspondence, Nucl. Phys. {\bf
B546} (1999) 96, hep-th/9804058.

\bibitem{symanzik0} K. Symanzik, Commun. Math. Phys. {\bf 23} (1971) 49.

\bibitem{ruehl1}
K. Lang and W. R\"uhl, Nucl.Phys. {\bf B377} (1992) 371.

\bibitem{ruehl2}
K. Lang and W. R\"uhl, Nucl. Phys. {\bf B400} (1993) 597;
Nucl. Phys. {\bf B402} (1993) 573.



\bibitem{tassos1}
A. C. Petkou,  ``Conserved currents, consistency relations
and operator product expansions in the conformally invariant $O(N)$
vector model for $2<d<4$'', Ann. Phys. {\bf 249} (1996) 180,
hep-th/9419093; ``Operator product expansions and consistency
relations in a $O(N)$ invariant fermionic CFT for $2<d<4$'',
Phys. Lett. {\bf  B389} (1996) 18, hep-th/9692954. 


\bibitem{tassos2}
H. Osborn and A. C. Petkou, ``Implications of conformal invariance for
field theories in general dimensions'', Ann. Phys. {\bf 231} (1994)
311, hep-th/9307010. 

\bibitem{dhoker}
E.~D'Hoker, S.D.~Mathur, A.~Matusis and L.~Rastelli, ``The operator
product expansion of {\cal N}=4 SYM and the 4-point functions of
supergravity'', hep-th/9911222.

\bibitem{freedman1}
D.Z.~Freedman, S.D.~Mathur, A.~Matusis and L.~Rastelli,
``Comments on 4-point functions in the CFT/AdS correspondence,"
Phys. Lett. {\bf B452} (1999) 61, hep-th/9808006;
E. D'Hoker and D. Z. Freedman, `General scalar
exchange in $AdS_{d+1}$', Nucl. Phys. {\bf B550} (1999) 261, hep--th/9811257;
E.~D'Hoker, D.Z. Freedman, S.D. Mathur, A. Matusis, L. Rastelli,
``Graviton and gauge boson propagators in $AdS_{d+1}$,"
Nucl. Phys. {\bf B562} (1999) 330,
hep-th/9902042;
E.~D'Hoker, D.Z.~Freedman, S.D.~Mathur, A.~Matusis and L.~Rastelli,
``Graviton exchange and complete 4-point functions in the AdS/CFT
correspondence,'' Nucl. Phys. {\bf B562} (1999) 353,
hep-th/9903196;
E.~D'Hoker, D.Z.~Freedman and L.~Rastelli,
``AdS/CFT 4-point functions: How to succeed at z-integrals without
really trying,'' Nucl. Phys. {\bf B562} (1999) 395,
hep-th/9905049.


\bibitem{liu}
H. Liu, `Scattering in Anti-de Sitter space and operator
product expansion', Phys. Rev. {\bf D60} 106005 (1999), hep--th/9811152.


\bibitem{tassos3}
A. C. Petkou and K. Skenderis, ``A non-renormalization theorem for
conformal anomalies'', Nucl. Phys. {\bf B561} (1999) 100, hep-th/9906030.

\bibitem{seiberg}
S. Lee, S. Minwalla, M. Rangamani and N. Seiberg, ``Three-point
functions of chiral primary operators in $d=4$, ${\cal N}=4$ SYM and
large $N$'', Adv. Theor. math. Phys. {\bf 2} (1998) 697, hep-th/9806074.

\bibitem{gleb1}
G. Arutyunov and S. Frolov, ``Some cubic coupling in type IIB
supergravity on AdS$_4\times$ S$^4$ and three-point functions in
SYM$_4$ for large $N$'', hep-th/9907085.

\bibitem{gleb2}
G. Arutyunov and S. Frolov, ``Scalar quartic couplings in type IIB
supergravity on AdS$_5\times$ S$^5$'', hep-th/9912210.

\bibitem{symanzik}
K. Symanzik, Lett. Nuovo Cim. {\bf 3} (1972), 734.


\bibitem{grad}
I. S. Gradsteyn and I. M. Ryshik, {\it Table of Integrals, Series and
  Products}, 5th Ed. Academic Press (1994).

\bibitem{HPR1}
L. Hoffmann, A. C. Petkou and W. R\"uhl, ``A note on the analyticity
of AdS scalar exchange graphs in the crossed channel'',
hep-th/0002025.

\bibitem{depp}
M. D'Eramo, G. Parisi and L. Peliti, Lett. Nuovo Cim. {\bf 2} (1971), 878.

\bibitem{parisi}
S. Ferrara, A. F. Grillo and G. Parisi, Lett. Nuovo Cim. {\bf 5}
(1972) 147; 
S. Ferrara, A. F. Grillo, R. Gatto and G. Parisi, Lett. Nuovo
Cim. {\bf 4} (1972) 115;
S. Ferrara and G. Parisi, Nucl. Phys. {\bf B42} (1972) 281.

\bibitem{koller}
K. Koller, Commun. Math. Phys. {\bf 40} (1975) 15.




\bibitem{klebanov}
I. Klebanov and E. Witten, ``AdS/CFT correspondence and symmetry breaking'',
Nucl. Phys. {\bf B556} (1999) 89, hep-th/9905104.

\bibitem{viswa2}
W. M\"uck and K. S. Viswanathan, ``Regular and irregular boundary
conditions in the AdS/CFT correspondence'', Phys. Rev. {\bf D60}
(1999) 081901, hep-th/9906155.

\bibitem{LeonRuehl}

T. Leonhard and W. R\" uhl, to appear.

\bibitem{todorov2}
V. K. Dobrev, G. Mack, V. B. Petkova, S. G. Petrova and I. T. Todorov,
Rep. in Math. Phys. {\bf 9} (1976) 219; {\it Harmonic analysis on the
$n$-dimensional conformal group and its applications to conformal
quantum field theory}, Springer-Verlag (1997).

\bibitem{freedman4}
E. D'Hoker, D. Z. Freedman and W. Skiba, ``Field theory tests for
correlators in the AdS/CFT correspondence'', Phys. Rev. {\bf D59}
(1999) 045008, hep-th/9807098.


\bibitem{slater}
L. J. Slater, {\it Generalized Hypergeometric Functions}, Cambridge
University Press (1966).



\bibitem{norlund}
A. E. N\/orlund, ``Hypergeometric functions'', Acta Math. {\bf
  94} (1955) 289.

\bibitem{herzog} 
C. P. Herzog, ``OPEs and 4-point functions if AdS/CFT
correspondence'', hep-th/0002039.

\bibitem{freedman5}
E. D'Hoker, D. Z. Freedman, S. D. Mathur, A. Matusis and L. Rastelli,
``Extremal correlators in the AdS/CFT correspondence'', hep-th/9908160.

\bibitem{gualtieri}
L. Gualtieri, ``Harmonic analysis and superconformal gauge theories in
three-dimensions from AdS/CFT correspondence'', hep-th/0002116. 

\end{thebibliography}
\end{document}